\RequirePackage{lineno}

\documentclass[aps,prc,twocolumn,superscriptaddress]{revtex4}

\usepackage{ulem}
\usepackage{epsfig}
\usepackage{graphicx}
\usepackage{fancyhdr}
\usepackage{pslatex}   
\usepackage{color}
\usepackage{amsmath, amsthm, amssymb} 
\usepackage{rotating}
\usepackage{multirow}

\usepackage{lineno}

\usepackage{ednmath0}

\newcommand {\Alm} {{\mbox{$A_{l,m}$}}}
\newcommand {\AlmQ} {{\mbox{$A_{l,m}\left(|\vec{q}|\right)$}}}

\newcommand {\pp} {\mbox{$p+p$}~}
\newcommand {\ppbar} {\mbox{$p+\bar{p}$}~}                                               
\newcommand {\epem} {\mbox{$e^{+}+e^{-}$}~}
\newcommand {\dAu} {\mbox{$d+\mathrm{Au}$}~}
\newcommand {\CuCu} {\mbox{$\rm Cu+Cu$}~}
\newcommand {\AuAu} {\mbox{$\rm Au+Au$}~}
\newcommand {\ApA} {\mbox{$\rm A+A$}~}

\newcommand {\roots} {\mbox{$\sqrt{s}$}}
\newcommand {\rootsTWO} {\mbox{$\sqrt{s}=200$~GeV}~}

\newcommand{\GeVc}{\mbox{$\mathrm{GeV}/c$}}
\newcommand{\cGeV}{\mbox{$c/\mathrm{GeV}$}}

\newcommand{\MeVc}{\mbox{$\mathrm{MeV}/c$}}

\begin{document}


\linenumberdisplaymath

\title {Pion femtoscopy in \pp collisions at \rootsTWO}

\affiliation{Argonne National Laboratory, Argonne, Illinois 60439, USA}
\affiliation{University of Birmingham, Birmingham, United Kingdom}
\affiliation{Brookhaven National Laboratory, Upton, New York 11973, USA}
\affiliation{University of California, Berkeley, California 94720, USA}
\affiliation{University of California, Davis, California 95616, USA}
\affiliation{University of California, Los Angeles, California 90095, USA}
\affiliation{Universidade Estadual de Campinas, Sao Paulo, Brazil}
\affiliation{University of Illinois at Chicago, Chicago, Illinois 60607, USA}
\affiliation{Creighton University, Omaha, Nebraska 68178, USA}
\affiliation{Czech Technical University in Prague, FNSPE, Prague, 115 19, Czech Republic}
\affiliation{Nuclear Physics Institute AS CR, 250 68 \v{R}e\v{z}/Prague, Czech Republic}
\affiliation{University of Frankfurt, Frankfurt, Germany}
\affiliation{Institute of Physics, Bhubaneswar 751005, India}
\affiliation{Indian Institute of Technology, Mumbai, India}
\affiliation{Indiana University, Bloomington, Indiana 47408, USA}
\affiliation{Alikhanov Institute for Theoretical and Experimental Physics, Moscow, Russia}
\affiliation{University of Jammu, Jammu 180001, India}
\affiliation{Joint Institute for Nuclear Research, Dubna, 141 980, Russia}
\affiliation{Kent State University, Kent, Ohio 44242, USA}
\affiliation{University of Kentucky, Lexington, Kentucky, 40506-0055, USA}
\affiliation{Institute of Modern Physics, Lanzhou, China}
\affiliation{Lawrence Berkeley National Laboratory, Berkeley, California 94720, USA}
\affiliation{Massachusetts Institute of Technology, Cambridge, MA 02139-4307, USA}
\affiliation{Max-Planck-Institut f\"ur Physik, Munich, Germany}
\affiliation{Michigan State University, East Lansing, Michigan 48824, USA}
\affiliation{Moscow Engineering Physics Institute, Moscow Russia}
\affiliation{City College of New York, New York City, New York 10031, USA}
\affiliation{NIKHEF and Utrecht University, Amsterdam, The Netherlands}
\affiliation{Ohio State University, Columbus, Ohio 43210, USA}
\affiliation{Old Dominion University, Norfolk, VA, 23529, USA}
\affiliation{Panjab University, Chandigarh 160014, India}
\affiliation{Pennsylvania State University, University Park, Pennsylvania 16802, USA}
\affiliation{Institute of High Energy Physics, Protvino, Russia}
\affiliation{Purdue University, West Lafayette, Indiana 47907, USA}
\affiliation{Pusan National University, Pusan, Republic of Korea}
\affiliation{University of Rajasthan, Jaipur 302004, India}
\affiliation{Rice University, Houston, Texas 77251, USA}
\affiliation{Universidade de Sao Paulo, Sao Paulo, Brazil}
\affiliation{University of Science \&amp; Technology of China, Hefei 230026, China}
\affiliation{Shandong University, Jinan, Shandong 250100, China}
\affiliation{Shanghai Institute of Applied Physics, Shanghai 201800, China}
\affiliation{SUBATECH, Nantes, France}
\affiliation{Texas A\&amp;M University, College Station, Texas 77843, USA}
\affiliation{University of Texas, Austin, Texas 78712, USA}
\affiliation{Tsinghua University, Beijing 100084, China}
\affiliation{United States Naval Academy, Annapolis, MD 21402, USA}
\affiliation{Valparaiso University, Valparaiso, Indiana 46383, USA}
\affiliation{Variable Energy Cyclotron Centre, Kolkata 700064, India}
\affiliation{Warsaw University of Technology, Warsaw, Poland}
\affiliation{University of Washington, Seattle, Washington 98195, USA}
\affiliation{Wayne State University, Detroit, Michigan 48201, USA}
\affiliation{Institute of Particle Physics, CCNU (HZNU), Wuhan 430079, China}
\affiliation{Yale University, New Haven, Connecticut 06520, USA}
\affiliation{University of Zagreb, Zagreb, HR-10002, Croatia}

\author{M.~M.~Aggarwal}\affiliation{Panjab University, Chandigarh 160014, India}
\author{Z.~Ahammed}\affiliation{Lawrence Berkeley National Laboratory, Berkeley, California 94720, USA}
\author{A.~V.~Alakhverdyants}\affiliation{Joint Institute for Nuclear Research, Dubna, 141 980, Russia}
\author{I.~Alekseev~~}\affiliation{Alikhanov Institute for Theoretical and Experimental Physics, Moscow, Russia}
\author{J.~Alford}\affiliation{Kent State University, Kent, Ohio 44242, USA}
\author{B.~D.~Anderson}\affiliation{Kent State University, Kent, Ohio 44242, USA}
\author{D.~Arkhipkin}\affiliation{Brookhaven National Laboratory, Upton, New York 11973, USA}
\author{G.~S.~Averichev}\affiliation{Joint Institute for Nuclear Research, Dubna, 141 980, Russia}
\author{J.~Balewski}\affiliation{Massachusetts Institute of Technology, Cambridge, MA 02139-4307, USA}
\author{L.~S.~Barnby}\affiliation{University of Birmingham, Birmingham, United Kingdom}
% ...

\collaboration{STAR Collaboration}\noaffiliation


\date{\today}

\begin{abstract}
The STAR Collaboration at RHIC has measured two-pion correlation functions from \pp collisions at \rootsTWO.
Spatial scales are extracted via a femtoscopic analysis of the correlations, though this analysis is complicated 
by the presence of strong non-femtoscopic effects.
Our results are put into the context of the world dataset of femtoscopy in hadron-hadron collisions. 
We present the first direct comparison of femtoscopy in \pp and heavy ion collisions, under identical analysis 
and detector conditions. 
\end{abstract}


\maketitle


\section{Introduction and Motivation}
\label{sec:intro}

The experimental program of the Relativistic Heavy Ion Collider (RHIC) at Brookhaven National Laboratory probes
Quantum Chromodynamics (QCD) from numerous directions.  The extraordinary flexibility of the machine permits
collisions between heavy and light ions at record energies (up to \rootsTWO), polarized and unpolarized protons,
and strongly asymmetric systems such as \dAu.  The proton collisions are the focus of an intense program exploring
the spin structure of the nucleon.  However, these collisions also serve as a critical ``baseline''
measurement for the heavy ion physics program that drove the construction of RHIC.

Studies of ultrarelativistic heavy ion collisions aim to explore the equation of state of strongly interacting matter.
The highly dynamic nature of the collisions, however, does not allow a purely statistical study of static matter as
one might perform in condensed matter physics, but rather requires a detailed understanding of the dynamics itself.
If a bulk, self-interacting system is formed (something that should not be assumed {\it a priori}),
the equation of state then plays the dynamic role of generating pressure gradients that drive the collective
expansion of the system.
Copious evidence~\cite{Adams:2005dq,Adcox:2004mh,Back:2004je,Arsene:2004fa} indicates that a self-interacting system
is, in fact, generated in these collisions.  
The dynamics of the bulk medium is reflected in the transverse momentum ($p_T$)
distribution~\cite{Schnedermann:1993ws,Retiere:2003kf} 
and momentum-space anisotropy (e.g. ``elliptic flow'')~\cite{Ollitrault:1992bk,Voloshin:2008dg}
of identified particles 
at low $p_T$.  These observables are well-described in a hydrodynamic scenario, in which a nearly perfect (i.e. very
low viscocity) fluid expands explosively under the action of pressure gradients induced by the collision~\cite{Kolb:2003dz}.

  Two-particle femtoscopy~\cite{Lednicky:2005af} (often called ``HBT'' analysis) measures the space-time substructure of the
emitting source at ``freeze-out,'' the point at which particles decouple from the system~\citep[e.g.][]{Lisa:2005dd}.
  Femtoscopic measurements play a special role in understanding bulk dynamics in heavy ion collisions, for several reasons.
  Firstly, collective flow generates characteristic space-momentum patterns at freezeout that are revealed~\cite{Lisa:2005dd}
in the momentum-dependence of pion ``HBT radii'' (discussed below), the transverse mass dependence of homogeneity
lengths~\cite{Akkelin:1995gh}, and non-identical particle correlations~\cite{Lednicky:2005af,Lednicky:1995vk}.
  Secondly, while a simultaneous description of particle-identified $p_T$ distributions, elliptic flow and femtoscopic
measurements is easily achieved in flow-dominated toy models~\citep[e.g.][]{Retiere:2003kf}, achieving the same level of agreement
in a realistic transport calculation is considerably more challenging.
  In particular, addressing this ``HBT puzzle''~\cite{Heinz:2002un} has led to a deeper understanding of 
the freezeout hypersurface,  
collectivity in the initial stage, 
and the equation of state.  
  Femtoscopic signals of long dynamical timescales expected for a system undergoing a first-order phase transition~\cite{Pratt:1986cc,Rischke:1996em},
have not been observed~\cite{Lisa:2005dd}, providing early evidence that the system at RHIC evolves from QGP to hadron gas via a crossover~\cite{Bekele:2007ee}.
  This sensitive and unique connection to important underlying physics has motivated a huge systematic study
of femtoscopic measurements in heavy ion collisions over the past quarter century~\cite{Lisa:2005dd}.

  HBT correlations from hadron (e.g. \pp) and lepton (e.g. \epem) collisions have been extensively studied in the high energy physics
community, as well~\cite{Kittel:2001zw,Alexander:2003ug,Chajecki:2009zg}, although
the theoretical interpretation of the results is less clear and not well developed.
  Until now, it has been impossible to quantitatively compare femtoscopic results from hadron-hadron collisions to
those from heavy ion collisions, due to divergent and often undocumented analysis techniques, detector acceptances
and fitting functions historically used in the high energy community~\cite{Chajecki:2009zg}.

  In this paper, we exploit the unique opportunity offered by the STAR/RHIC experiment, to make the first direct
comparison and quantitative connection between femtoscopy in proton-proton and heavy ion collisions.
  Systematic complications in comparing these collisions are greatly reduced by using an identical
detector and reconstruction software, collision energies, and analysis techniques (e.g. event mixing~\cite{Kopylov:1974th},
see below).
  We observe and discuss the importance of non-femtoscopic correlations in the analysis of small systems, and
put our femtoscopic results for \pp collisions into the context both of heavy ion collisions and (as much as possible)
of previous high-energy measurements on hadron-hadron and \epem collisions.
These results may play a role in understanding
the physics behind the space-momentum
correlations in these collisions, in the same way that comparison of \pp and heavy ion collision results in the high-$p_T$
sector is crucial for understanding the physics of partonic energy loss~\cite{Adams:2005dq,Adcox:2004mh,Back:2004je,Arsene:2004fa,Jacobs:2004qv}.
  Our direct comparison also serves as a model and baseline for similar comparisons soon to be possible at higher
energies at the Large Hadron Collider.

The paper is organized as follows.  In Section~\ref{sec:formalism}, we discuss the construction of the
correlation function and the forms used to parameterize it.
Section~\ref{sec:details} discusses details of the analysis, and the results are presented in Section~\ref{sec:results}.
In Section~\ref{sec:WorldSystematics}, we put these results in the  context of previous measurements in \AuAu and $p+p(\bar{p})$ collisions.
We discuss the similarity between the systematics of HBT radii in heavy ion and particle collisions in Section~\ref{sec:discussion}
and summarize in Section~\ref{sec:summary}.

\vspace*{-2mm}

\section{Two-particle correlation function}
\label{sec:formalism}

The two-particle correlation function is generally defined as the ratio of the probability
of the simultaneous measurement of two particles with momenta $p_1$ and $p_2$,
to the product of single-particle probabilities, 
\begin{linenomath*}
\begin{equation}
\label{eq:c2}
C(\vec{p}_1,\vec{p}_2) \equiv \frac{P(\vec{p}_1,\vec{p}_2)}{P(\vec{p}_1)P(\vec{p}_2)}.
\end{equation}
\end{linenomath*}
In practice, one usually studies the quantity
\begin{linenomath*}
\begin{equation}
\label{eq:CFexp}
C_{\vec{P}}\left(\vec{q}\right)=\frac{A_{\vec{P}}\left(\vec{q}\right)}{B_{\vec{P}}\left(\vec{q}\right)} ,
\end{equation}       
\end{linenomath*}
where $\vec{q}\equiv\vec{p_1}-\vec{p_2}$ is the relative momentum.
$A(\vec{q})$ is the distribution of the pairs from the same event, and $B(\vec{q})$ 
is the reference  (or ``background'') distribution.
$B$ contains all single-particle effects, including detector acceptance and efficiency,
and is usually calculated with an event-mixing technique~\cite{Kopylov:1974th,Lisa:2005dd}.
The explicit label $\vec{P}$ ($\equiv\left(\vec{p_1}+\vec{p_2}\right)/2$) emphasizes that separate correlation
functions are constructed and fitted (see below) as a function of $\vec{q}$, for different
selections of the total momentum $\vec{P}$; following convention, we drop the explicit subscript below.
Sometimes the measured ratio is normalized to unity at large values of $|\vec{q}|$; we include
the normalization in the fit.

In older or statistics-challenged experiments, the correlation function is sometimes
constructed in the one-dimensional quantity $Q_{\rm inv}\equiv\sqrt{\left(\vec{p_1} - \vec{p_2}\right)^2-\left(E_1 - E_2\right)^2}$
or two-dimensional variants (see below).
More commonly in recent experiments, it is constructed in three dimensions in the so-called 
the ``out-side-long'' coordinate 
system~\cite{Podgoretsky:1982xu,Pratt:1984su,Bertsch:1988db}.  In this system, the ``out'' direction is that of the
pair transverse momentum, the ``long'' direction is parallel to the beam, and the ``side'' direction
is orthogonal to these two.  We will use the subscripts ``$o$,'' ``$l$'' and ``$s$'' to indicate
quantities in these directions.

It has been suggested~\cite{Danielewicz:2005qh,Danielewicz:2006hi,Chajecki:2008vg} to construct the
three-dimensional
correlation function using spherical coordinates
\begin{linenomath*}
\begin{equation}
\label{eq:angleDefinitions}
q_{o} = |\vec{q}|\sin{\theta}\cos{\phi} , \qquad
q_{s} = |\vec{q}|\sin{\theta}\sin{\phi} , \qquad
q_{l} = |\vec{q}|\cos{\theta} .
\end{equation}
\end{linenomath*}
This aids in making a direct comparison to the spatial separation distribution through
imaging techniques and provides an efficient way to visualize
the full three-dimensional structure of $C\left(\vec{q}\right)$.
The more traditional
``Cartesian projections'' in the ``$o$,'' ``$s$'' and ``$l$'' directions integrate
over most of the three-dimensional structure, especially at large relative momentum~\cite{Lisa:2005dd,Chajecki:2008vg}.

Below, we will present data in the form of the spherical harmonic decomposition coefficients, which
depend explicitly on $|\vec{q}|$ as
\begin{linenomath*}
\begin{equation}
\label{eq:Alm}
\AlmQ \equiv \frac{1}{\sqrt{4\pi}} \int d\phi d(\cos\theta)  C\left(|\vec{q}|,\theta,\phi \right) Y_{l,m}\left(\theta,\phi\right).
\end{equation}
\end{linenomath*}
The coefficient $A_{00}\left(|\vec{q}|\right)$ represents the overall angle-integrated strength of
the correlation.  $A_{20}\left(|\vec{q}|\right)$ and $A_{22}\left(|\vec{q}|\right)$ are the quadrupole moments of $C$
at a particular value of $|\vec{q}|$.  In particular, $A_{22}$ quantifies the second-order oscillation  
around the ``long'' direction;
in the simplest HBT analysis, this term reflects non-identical values of the $R_{o}$ and $R_{s}$ HBT radii (c.f. below).  Coefficients with odd $l$
represent a dipole moment of the correlation function and correspond to a ``shift'' in the average position of the first particle in a pair, relative to
the second~\cite{Danielewicz:2005qh,Danielewicz:2006hi,Chajecki:2008vg}.  In the present case of identical particles, the labels ``first'' and ``second''
become meaningless, and odd-$l$ terms vanish by symmetry.  Likewise, for the present case, odd-$m$ terms, and all imaginary components vanish as well.
See Appendix B of~\cite{Chajecki:2008vg} for a full discussion of symmetries.

In heavy ion collisions, it is usually assumed that all of the correlations between identical pions at low relative
momentum are due to femtoscopic effects, i.e. quantum statistics and final-state interactions~\cite{Lisa:2005dd}.
At large $|\vec{q}|$, femtoscopic effects vanish~\citep[e.g.][]{Lisa:2005dd}.
Thus, in the absence of other correlations, $C\left(\vec{q}\right)$ must approach a constant value independent of
the magnitude and direction of $\vec{q}$; equivalently, $\AlmQ$ must vanish at large $|\vec{q}|$ for $l\neq 0$.

However, in elementary particle collisions additional structure at large relative momentum ($|\vec{q}|\gtrsim 400$~\MeVc) has been
observed~\citep[e.g.][]{Avery:1985qb,Agababyan:1996rg,collaboration:2007he,Bailly:1988zb,Uribe:1993tr,Chajecki:2009zg}.
Usually this structure is parameterized in terms of a function $\Omega\left(\vec{q}\right)$
that contributes in addition to the femtoscopic component $C_F\left(\vec{q}\right)$.
Explicitly including the normalization parameter $\mathcal{N}$, then, we
will fit our measured correlation functions with the form
\begin{linenomath*}
\begin{equation}
\label{eq:factorization}
C\left(\vec{q}\right) = \mathcal{N} \cdot C_F\left(\vec{q}\right) \cdot \Omega\left(\vec{q}\right) .
\end{equation}
\end{linenomath*}
Below, we discuss separately various parameterizations of the femtoscopic and non-femtoscopic components,
which we use in order to connect with previous measurements.
A historical discussion of these forms may be found in~\cite{Chajecki:2009zg}.

We use a maximum-likelihood fit to the correlation functions, though chi-square minimization yields almost
identical results, and we give the $\chi^2$ values for all fits below.  As we shall see, none of the functional
forms perfectly fits the data.  However, the characteristic scales of the source can be extracted and compared
with identical fits to previous data.

\vspace*{-4mm}

\subsection{Femtoscopic correlations}
\label{sec:femto}
Femtoscopic correlations between identical pions are dominated by Bose-Einstein symmetrization and Coulomb final state effects
in the two-pion wavefunction~\cite{Lisa:2005dd}.

In all parameterizations, the overall strength of the femtoscopic correlation is characterized by a 
parameter $\lambda$~\cite{Lisa:2005dd}.
Historically called the ``chaoticity'' parameter, it generally accounts for particle identification efficiency,
long-lived decays, and long-range tails in the separation distribution~\cite{Lednicky:1979ig}. 

In the simplest case, the Bose-Einstein correlations are often parameterized by
a Gaussian,
\begin{linenomath*}
\begin{equation}                                                                                         
\label{eq:CFqinv}                                                                                        
C_F(Q_{\rm inv})=  1 + \lambda e^{-Q^2_{\rm inv} R^{2}_{\rm inv}},                                                   
\end{equation}                                                                                           
\end{linenomath*}
where $R_{\rm inv}$ is a one dimensional ``HBT radius.''

Kopylov and Podgoretskii~\cite{Kopylov:1972qw} introduced an alternative, two-dimensional parameterization
\begin{linenomath*}
\begin{equation}                                                                                         
\label{eq:CFcp}                                                                                          
C_F(q_T,q_0)=  1 + \lambda \left[ \frac{2 J_1 \left( q_T R_B \right) }{q_T R_B}\right]^2 \left(1 +  q^2_0 \tau^2 \right)^{-1},
\end{equation}                                                                                           
\end{linenomath*}
where $q_T$ is the component of $\vec{q}$ orthogonal to $\vec{P}$, 
$q_0=E_1-E_2$, $R_B$ and $\tau$ are the size and decay constants of                      
a spherical emitting source,                                                                             
and $J_1$ is the first order Bessel function.                                            
This is similar to another common historical parameterization~\citep[e.g.][]{Alexopoulos:1992iv} characterizing
the source with a spatial and temporal scale
\begin{linenomath*}
\begin{equation}                                                                                         
\label{eq:CFgauss}                                                                                        
C_F(q,q_0)=  1 + \lambda e^{-q_T^2 R^{2}_{G}-q_0^2 \tau^2} .                                                   
\end{equation}                                                                                           
\end{linenomath*}

Simple numerical studies
show that $R_{G}$ from                                                                                   
Eq.~\ref{eq:CFgauss} is approximately 
half as large as
$R_{B}$ obtained from Eq.~\ref{eq:CFcp}~\cite{Boal:1990yh,Alexopoulos:1992iv,Chajecki:2009zg}.

With sufficient statistics, a three-dimensional correlation function may be measured.
We calculate the relative momentum in the longitudinally co-moving system (LCMS), in which
the total longitudinal momentum of the pair, $p_{l,1}+p_{l,2}$, vanishes~\cite{Pratt:1990zq}.  For heavy
ion and hadron-hadron collisions, this ``longitudinal'' direction $\hat{l}$ is taken to be the beam axis~\cite{Lisa:2005dd};
for \epem ~collisions, the thrust axis is used.  

For a Gaussian emission source, femtoscopic correlations due only to Bose-Einstein symmetrization
are given by~\citep[e.g.][]{Lisa:2005dd}
\begin{linenomath*}
\begin{equation}                                                                                         
\label{eq:cf3D_noCoulomb}                                                                                          
C_F(q_o,q_s,q_l)=  1 + \lambda e^{-q^2_o R^2_o - q^2_s R^2_s - q^2_l R^2_l},                             
\end{equation}                                                                                           
\end{linenomath*}
where  $R_o$, $R_s$ and $R_l$ are the spatial scales of the source.

While older papers sometimes ignored the Coulomb final-state interaction between the charged pions~\cite{Chajecki:2009zg}, it is
usually included by using the Bowler-Sinyukov~\cite{Bowler:1991vx,Sinyukov:1998fc} functional form
\begin{linenomath*}
\begin{equation}                                                                                         
\label{eq:CFqinvWithCoulomb}                                                                                        
C_F(Q_{\rm inv})=  \left(1 - \lambda\right) + \lambda K_{\rm coul}\left(Q_{\rm inv}\right) \left( 1 + e^{-Q^2_{\rm inv} R^{2}_{\rm inv}} \right),
\end{equation}                                                                                           
\end{linenomath*}
and in 3D, 
\begin{linenomath*}
\begin{eqnarray}
\label{eq:cf3D}                                                                                          
C_F(q_o,q_s,q_l) &=&  \left(1 - \lambda\right) + \lambda K_{\rm coul}\left(Q_{\rm inv}\right) \nonumber \\
&& \times \left(1 +  e^{-q^2_o R^2_o - q^2_s R^2_s - q^2_l R^2_l} \right).
\end{eqnarray}
\end{linenomath*}
Here, $K_{\rm coul}$ is the squared Coulomb wavefunction integrated over the source emission points
  and over the angles of the relative momentum vector in the pair rest frame.

\subsection{Non-femtoscopic correlations}
\label{sec:nonfemto}

In the absence of non-femtoscopic effects, one of the forms for $C_F\left(\vec{q}\right)$ from Section~\ref{sec:femto} is
fitted to the measured correlation function; i.e. $\Omega=1$ in Equation~\ref{eq:factorization}.  Such a ``standard fit''
works well in the high-multiplicity environment of heavy ion collisions~\cite{Lisa:2005dd}.  In hadron-hadron or \epem collisions,
however, it does not describe the measured correlation function well, especially as $|q|$ increases.
Most authors attribute the non-femtoscopic structure to momentum conservation effects in these small systems.
  While this large-$|q|$ behavior
is sometimes simply ignored,
 it is usually included in the fit either through ad-hoc~\cite{Agababyan:1996rg} or physically-motivated~\cite{Chajecki:2008vg}
terms.

In this paper, we will use three selected parameterizations of the non-femtoscopic correlations and
study their effects on the femtoscopic parameters obtained from the fit to experimental correlation 
functions. 
The first formula assumes that the non-femtoscopic contribution can be parameterized by a first-order
polynomial in $\vec{q}$-components (used e.g.  in \cite{Buskulic:1994ny,Achard:2001za,AguilarBenitez:1991ri,Abreu:1994wd,Abreu:1992gj}).
Respectively, the one- and three-dimensional forms used in the literature are
\begin{linenomath*}
\begin{equation}
\label{eq:dQ1D}
\Omega(q) = 1 + \delta q 
\end{equation}
\end{linenomath*}
and 
\begin{linenomath*}
\begin{equation}
\label{eq:dQ3D}
\Omega(\vec{q}) = \Omega(q_o, q_s, q_l) = 1 + \delta_o q_{o} + \delta_s q_{s} + \delta_l q_{l} .
\end{equation}
\end{linenomath*}
For simplicity, we will use the name  ``$\delta-q$ fit'' when we fit Eq.~\ref{eq:dQ1D} or~\ref{eq:dQ3D} to one- or three-dimensional correlation functions.

Another form~\cite{Chajecki:2005iv} assumes that
non-femtoscopic correlations contribute $|\vec{q}|$-independent values to the $l=2$ moments in Equation~\ref{eq:Alm}.
In terms of the fitting parameters $\zeta$ and $\beta$, 
\begin{linenomath*}
\begin{eqnarray}
\label{eq:ZetaBeta}
\Omega\left(|\vec{q}|, \cos\theta, \phi\right) = \Omega\left(\cos\theta, \phi\right) = \nonumber \\
 1 + 2\sqrt{\pi}\left(\beta Y_{2,0}\left(\cos\theta, \phi\right) + 2\zeta{\rm Re}\left[Y_{2,2}\left(\cos\theta, \phi\right)\right]\right) = \nonumber \\
 1 + \beta \sqrt{\frac{5}{4}} (3\cos^2 \theta -1)  + \zeta  \sqrt{\frac{15}{2}} \sin^2 \theta \cos 2\phi .
\label{eq:PBcfBaseline}                           
\end{eqnarray}
\end{linenomath*}
For simplicity, fits using this form for the non-femtoscopic effects will be referred
to as ``$\zeta-\beta$ fits.''

These two forms (as well as others that can be found in literature~\cite{Chajecki:2009zg})
  are purely empirical, motivated essentially by the shape of the observed correlation function itself.
While most authors attribute these effects primarily to momentum conservation in these low-multiplicity
  systems, the parameters and functional forms themselves cannot be directly connected to this or any physical mechanism.
One may identify two dangers of using an ad-hoc form to quantify non-femtoscopic contributions to $C\left(\vec{q}\right)$.
Firstly, while they describe (by construction) the correlation function well at large
  $|\vec{q}|$, for which femtoscopic contributions vanish, there is no way to constrain their behaviour
  at low $|\vec{q}|$ where both femtoscopic and (presumably) non-femtoscopic correlations exist.
Even simple effects like momentum conservation give rise to non-femtoscopic correlations that
  vary non-trivially even at low $|\vec{q}|$.  Misrepresenting the non-femtoscopic contribution
  in $\Omega\left(\vec{q}\right)$ can therefore distort the femtoscopic radius parameters in $C_F\left(\vec{q}\right)$, especially considering the small radius values in \pp collisions.
Secondly, there is no way to estimate whether the best-fit parameter values in an ad-hoc functional form
  are physically ``reasonable.'' 

If the non-femtoscopic correlations are in fact dominated by energy and momentum conservation, as is
  usually supposed, one may derive an analytic functional form for $\Omega$.
In particular, the multiparticle phase space constraints for a system of $N$ particles project onto the
  two-particle space as~\cite{Chajecki:2008vg}
\begin{linenomath*}
\begin{align}
\label{eq:EMCICs}
\Omega\left(p_1,p_2\right) =& 1
   - M_1\cdot\overline{\left\{\vec{p}_{1,T}\cdot\vec{p}_{2,T}\right\}}
   - M_2\cdot\overline{\left\{p_{1,z}\cdot p_{2,z}\right\}} \\
   &- M_3\cdot\overline{\left\{E_1 \cdot E_2\right\}}
   + M_4\cdot\overline{\left\{E_1 + E_2\right\}}
   - \frac{M_4^2}{M_3} , \qquad \nonumber
\end{align}
\end{linenomath*}
where
\begin{linenomath*}
\begin{alignat}{2}
\label{eq:fitParameters}
M_1&\equiv\frac{2}{N\langle p_T^2 \rangle},                            \quad  &M_2&\equiv\frac{1}{N\langle p_z^2 \rangle}  \nonumber \\
M_3&\equiv\frac{1}{N\left(\langle E^2\rangle-\langle E\rangle^2\right)},\quad &M_4&\equiv\frac{\langle E\rangle}{N\left(\langle E^2\rangle-\langle E\rangle^2\right)} .
\end{alignat}
\end{linenomath*}
 
The notation $\overline{\left\{X\right\}}$ in Equation~\ref{eq:EMCICs} is used to indicate that 
  $X$ is the average of a two-particle quantity which depends on $p_1$ and $p_2$ (or $\vec{q}$, etc).
In particular,
\begin{linenomath*}
\begin{equation}
\overline{\left\{X\right\}}\left(\vec{q}\right) \equiv
\frac{\int d^3\vec{p}_1 \int d^3\vec{p}_2 P\left(\vec{p}_1\right) P\left(\vec{p}_2\right) X \delta\left(\vec{q}-\left(\vec{p}_1-\vec{p}_2\right)\right)}{\int d^3\vec{p}_1 \int d^3\vec{p}_2 P\left(\vec{p}_1\right) P\left(\vec{p}_2\right) \delta\left(\vec{q}-\left(\vec{p}_1-\vec{p}_2\right)\right)} ,
\end{equation}
\end{linenomath*}
where $P$ represents the single-particle probability first seen in Equation~\ref{eq:c2}.

In practice, this means generating histograms in addition to $A\left(\vec{q}\right)$ 
  and $B\left(\vec{q}\right)$ (c.f. Equation~\ref{eq:CFexp}) as one loops over mixed pairs of particles $i$ and $j$ in the data analysis.
For example
\begin{linenomath*}
\begin{equation}
\overline{\left\{\vec{p}_{1,T}\cdot\vec{p}_{2,T}\right\}}\left(\vec{q}\right) =
\frac{\left(\sum_{i,j}\vec{p}_{i,T}\cdot\vec{p}_{j,T}\right)\left(\vec{q}\right)}{B\left(\vec{q}\right)} ,
\end{equation}
\end{linenomath*}
where the sum in the numerator runs over all pairs in all events.

In Equation~\ref{eq:EMCICs}, the four fit parameters $M_i$ are 
directly related to five physical quantities, ($N$ - the number of particles,
$\langle p_T^2 \rangle$, $\langle p_z^2 \rangle$, $\langle E^2 \rangle$, $\langle E \rangle$)  through Eq.~\ref{eq:fitParameters}.
Assuming that 
\begin{linenomath*}
\begin{equation}
\label{eq:Esq}
\langle E^2\rangle \approx \langle p_T^2 \rangle + \langle p_z^2 \rangle + m_{*}^2,
\end{equation}
\end{linenomath*}
where $m_{*}$ is the mass of a typical particle in the system (for our pion-dominated system, $m_{*}\approx m_\pi$),
then one may solve for the physical parameters.  For example,
\begin{linenomath*}
\begin{equation}
\label{eq:Nmin}
N \approx \frac{ M_{1}^{-1} + M_{2}^{-1} - M_{3}^{-1}}{ \left(\frac{M_4}{M_3}\right)^2 - m_{*}^2}.
\end{equation} 
\end{linenomath*}

Since we cannot know exactly the values of $\langle E^2\rangle$ etc, that characterize the underlying distribution
  in these collisions, we treat the $M_i$ as free parameters in our fits, and then consider whether their
  values are mutually compatible and physical.
For a more complete discussion, see~\cite{Chajecki:2008vg,Chajecki:2008yi}.
 
In~\cite{Chajecki:2008vg}, the correlations leading to Equation~\ref{eq:EMCICs} were called ``EMCICs''
(short for Energy and Momentum Conservation-Induced Correlations); we will refer to fits using this
function with this acronym, in our figures.

\subsection{Parameter counting}

As mentioned, we will be employing a number of different fitting functions, each of which contains
  several parameters.
It is appropriate at this point to briefly take stock.

In essentially all modern HBT analyses, on the order of 5-6 parameters quantify the femtoscopic
  correlations.  For the common Gaussian fit (equation~\ref{eq:cf3D}), one has three ``HBT radii,'' the chaoticity
  parameter, and the normalization $\mathcal{N}$.
Recent ``imaging'' fits approximate the two-particle emission zone as a sum of spline functions, the weights
  of which are the parameters~\cite{Brown:2005ze}; the number of splines (hence weights) used is $\sim 5$.
Other fits (e.g. double Gaussian, exponential-plus-Gaussian)~\cite{Kittel:2001zw,Kittel:2005fu} contain a similar number of femtoscopic parameters.
In all cases, a distinct set of parameters is extracted for each selection of $\vec{P}$ (c.f. equation~\ref{eq:CFexp}
  and surrounding discussion).

Accounting for the non-femtoscopic correlations inevitably increases the total number of fit parameters.
The ``$\zeta-\beta$'' functional form (eq.~\ref{eq:ZetaBeta}) involves two parameters, the ``$\delta-q$'' form (eq.~\ref{eq:dQ3D}) 
  three, and the EMCIC form (eq.~\ref{eq:EMCICs}) four.
However, it is important to keep in mind that using the $\zeta-\beta$ ($\delta-q$) form means 2 (3) additional
  parameters {\it for each selection} of $\vec{P}$ when forming the correlation functions.
On the other hand, the four EMCIC parameters cannot depend on $\vec{P}$.
Therefore, when fitting $C_{\vec{P}}\left(\vec{q}\right)$ for four selections of $\vec{P}$, use of the $\zeta-\beta$, $\delta-q$
  and EMCIC forms increases the total number of parameters by 8, 12 and 4, respectively.

\vspace*{-4mm}

\section{Analysis details}
\label{sec:details}

As mentioned in Section~\ref{sec:intro}, there is significant advantage in analyzing \pp collisions
  in the same way that heavy ion collisions are analyzed.
Therefore, the results discussed in this paper are produced with the same techniques and acceptance cuts
  as have been used for previous pion femtoscopy studies by STAR~\cite{Adler:2001zd,Adams:2003ra,Adams:2004yc,Abelev:2009tp}.
Here we discuss some of the main points; full systematic studies of cuts and techniques can be found
  in~\cite{Adams:2004yc}.

The primary sub-detector used in this analysis to reconstruct particles
  is the Time Projection Chamber (TPC)~\cite{Anderson:2003ur}. 
Pions could be identified up to a momentum of 800~\MeVc~by correlating their
  momentum and specific ionization loss ($dE/dx$) in the TPC gas.
A particle was considered to be a pion if its $dE/dx$ value for a given momentum
  was within two sigma of the Bichsel expectation~\cite{Bichsel:2006cs} (an improvement on the Bethe-Bloch formula~\cite{Yao:2006px} for thin materials)
 for a pion, and more than two sigma from the
  expectations for electrons, kaons and protons.
By varying the cuts on energy loss to allow more or less contamination from kaons or electrons, we
  estimate that impurities in the pion sample lead to an uncertainty in the femtoscopic scale parameters (e.g. HBT radii) of only about 1\%.
Particles were considered for analysis if their reconstructed tracks produced hits on at least 10 of the 45 padrows, and
  their distance of closest approach (DCA) to the primary vertex was less than 3~cm.
The lower momentum cut of 120~\MeVc ~is imposed by the TPC acceptance and the magnetic field. 
Only tracks at midrapidity ($|y|<0.5$) were included in the femtoscopic analysis.

Events were recorded based on a coincidence trigger of two Beam-Beam Counters (BBCs), annular scintillator
detectors located $\pm 3.5$~m from the interaction region and covering pseudorapidity range $3.3<|\eta|<5.0$.
Events were selected for analysis if the primary collision vertex was within 30~cm of the center of the TPC.
The further requirement that events include at least two like-sign pions increases the average charged
  particle multiplicity with $|\eta|<0.5$ from 3.0 (without the requirement) to 4.25.
Since particle {\it pairs} enter into the correlation function, the effective average multiplicity is
  higher; in particular, the pair-weighted charged-particle multiplicity at midrapidity is about 6.0.
After event cuts, about 5 million minimum bias events from \pp collisions at \rootsTWO were used.

Two-track effects, such as splitting (one particle reconstructed as two tracks) and merging 
  (two particles reconstructed as one track) were treated identically as has been done in
  STAR analyses of \AuAu collisions~\cite{Adams:2004yc}.
Both effects can affect the shape of $C\left(\vec{q}\right)$ at very low $|\vec{q}|\lesssim 20$~\MeVc,
  regardless of the colliding system.
However, their effect on the extracted sizes in \pp collisions turns out to be smaller than
  statistical errors, due to the fact that small ($\sim 1$~fm) sources lead to large ($\sim 200$~\MeVc) 
  femtoscopic structures in the correlation function.

The analysis presented in this paper was done for four bins 
  in average transverse momentum $k_T$~($\equiv\frac{1}{2}|\left(\vec{p}_{T,1}+\vec{p}_{T,2}\right)|$):
  150-250, 250-350, 350-450 and 450-600~\MeVc.
The systematic errors on femtoscopic radii due to the fit range, particle mis-identification, two-track effects and 
  the Coulomb radius (used to calculate $K_{\rm coul}$ in Eqs.~\ref{eq:CFqinvWithCoulomb} and~\ref{eq:cf3D}) 
  are estimated to be about 10\%, similar to previous studies~\cite{Adams:2004yc}.

\vspace*{-4mm}

\section{Results}
\label{sec:results}

In this section, we present the correlation functions and fits to them, using the various functional forms
discussed in Section~\ref{sec:formalism}.  The $m_T$ and multiplicity dependence of femtoscopic radii from
these fits are compared here, and put into the broader context of data from heavy ion and particle collisions
in the next section.

Figure~\ref{fig:CartesianProjections} shows the two-pion correlation function
  for minimum-bias \pp collisions for $0.35<k_T<0.45$~\GeVc.
The three-dimensional data is represented with the traditional one-dimensional Cartesian projections~\cite{Lisa:2005dd}.
For the projection on $q_o$, integration in $q_s$ and $q_l$ was done over the range $[0.00, 0.12]$~\GeVc. 
As discussed in Section~\ref{sec:formalism} and in more detail in~\cite{Chajecki:2008vg}, the full structure 
  of the correlation function is best seen in the spherical harmonic decomposition, shown in Figs.~\ref{fig:SHDkt1}-\ref{fig:SHDkt4}.

In what follows, we discuss systematics of fits to the correlation function, with particular attention to the femtoscopic parameters.
It is important to keep in mind that the fits are performed on the full three-dimensional correlation function $C\left(\vec{q}\right)$.
The choice to plot the data and fits as spherical harmonic coefficients $A_{lm}$ or as Cartesian projections along the ``out,'' ``side''
  and ``long'' directions is based on the desire to present results in the traditional format (projections) or in a representation
  more sensitive to the three-dimensional structure of the data~\cite{Chajecki:2008vg}.
In particular, the data and fits shown in Fig.~\ref{fig:CartesianProjections}, for $k_T$=0.35-0.45~\GeVc, are the same as those shown in Fig.~\ref{fig:SHDkt3}.

\vspace*{-4mm}

\subsection{Transverse mass dependence of 3D femtoscopic radii}
\label{sec:mTdep3D}
\begin{table*}[ht!]
\begin{tabular}{|c||c|c|c|c|c|}
\hline
$k_T$ [\GeVc] & $R_{o}$ [fm] & $R_{s}$ [fm] & $R_{l}$ [fm] & $\lambda$ & $\chi^2/{\rm ndf}$  \\
\hline
$[0.15,0.25]$ &  $0.84 \pm 0.02$& $0.89 \pm 0.01$& $1.53 \pm 0.02$& $0.422 \pm 0.004$ & 2012 / 85 \\
$[0.25,0.35]$ &  $0.81 \pm 0.02$& $0.88 \pm 0.01$& $1.45 \pm 0.02$& $0.422 \pm 0.005$ & 1852 / 85 \\
$[0.35,0.45]$ &  $0.71 \pm 0.02$& $0.82 \pm 0.02$& $1.31 \pm 0.02$& $0.433 \pm 0.007$ & 941  / 85 \\
$[0.45,0.60]$ &  $0.68 \pm 0.02$& $0.68 \pm 0.01$& $1.05 \pm 0.02$& $0.515 \pm 0.009$ & 278  / 85 \\
\hline
\end{tabular}
\caption{Fit results from a fit to data from \pp collisions at \roots = 200 GeV using Eq.~\ref{eq:cf3D} to parameterize the femtoscopic correlations (``standard fit''). \label{tab:3DfitStandard}}
\end{table*}
\begin{table*}
\begin{tabular}{|c||c|c|c|c|c|c|c|c|}
\hline
$k_T$ [\GeVc] & $R_{o}$ [fm] & $R_{s}$ [fm] & $R_{l}$ [fm] & $\lambda$ & $\delta_{o}$ & $\delta_{s}$ & $\delta_{l}$ & $\chi^2/{\rm ndf}$  \\
\hline
$[0.15,0.25]$ &  $1.30 \pm 0.03$& $1.05 \pm 0.03$& $1.92 \pm 0.05$& $0.295 \pm 0.004$ & $ 0.0027 \pm 0.0026 $ & $ -0.1673 \pm 0.0052 $ & $ -0.2327 \pm 0.0078 $ & 471 / 82\\
$[0.25,0.35]$ &  $1.21 \pm 0.03$& $1.05 \pm 0.03$& $1.67 \pm 0.05$& $0.381 \pm 0.005$ & $ 0.0201 \pm 0.0054 $ & $ -0.1422 \pm 0.0051 $ & $ -0.2949 \pm 0.0081 $ & 261 / 82\\
$[0.35,0.45]$ &  $1.10 \pm 0.03$& $0.94 \pm 0.03$& $1.37 \pm 0.05$& $0.433 \pm 0.007$ & $ 0.0457 \pm 0.0059 $ & $ -0.0902 \pm 0.0053 $ & $ -0.2273 \pm 0.0090 $ & 251 / 82\\    
$[0.45,0.60]$ &  $0.93 \pm 0.03$& $0.82 \pm 0.03$& $1.17 \pm 0.05$& $0.480 \pm 0.009$ & $ 0.0404 \pm 0.0085 $ & $ -0.0476 \pm 0.0093 $ & $ -0.1469 \pm 0.0104 $ & 189 / 82\\   
\hline
\end{tabular}
\caption{Fit results from a fit to data from \pp collisions at \roots = 200 GeV using Eq.~\ref{eq:cf3D} 
to parameterize the femtoscopic correlations and Eq.~\ref{eq:dQ3D} for non-femtoscopic ones 
(``$\delta-q$ fit''). \label{tab:3DfitDeltaQ}}
\end{table*}
\begin{table*}
 \begin{tabular}{|c||c|c|c|c|c|c|c|}
\hline
$k_T$ [\GeVc]  & $R_{o}$ [fm] & $R_{s}$ [fm] & $R_{l}$ [fm] & $\lambda$  & $\zeta$  & $\beta$ & $\chi^2/{\rm ndf}$ \\
\hline
$[0.15,0.25]$ &  $1.24 \pm 0.04$& $0.92 \pm 0.03$& $1.71 \pm 0.04$& $0.392 \pm 0.008$ & $ 0.0169 \pm 0.0021 $ & $ -0.0113 \pm 0.0019 $ & 1720 / 83 \\
$[0.25,0.35]$ &  $1.14 \pm 0.05$& $0.89 \pm 0.04$& $1.37 \pm 0.08$& $0.378 \pm 0.006$ & $ 0.0193 \pm 0.0034 $ & $ -0.0284 \pm 0.0031 $ & 823  / 83 \\
$[0.35,0.45]$ &  $1.02 \pm 0.04$& $0.81 \pm 0.05$& $1.20 \pm 0.07$& $0.434 \pm 0.008$ & $ 0.0178 \pm 0.0029 $ & $ -0.0289 \pm 0.0032 $ & 313  / 83 \\ 
$[0.45,0.60]$ &  $0.89 \pm 0.04$& $0.71 \pm 0.05$& $1.09 \pm 0.06$& $0.492 \pm 0.009$ & $ 0.0114 \pm 0.0023 $ & $ -0.0301 \pm 0.0041 $ & 190  / 83 \\       
\hline
\end{tabular}
\caption{Fit results from a fit to data from \pp collisions at \roots = 200 GeV using Eq.~\ref{eq:cf3D} 
to parameterize the femtoscopic correlations and Eq.~\ref{eq:ZetaBeta} for non-femtoscopic ones 
(``$\zeta-\beta$ fit''). \label{tab:3DfitZetaBeta}}
\end{table*}
\begin{table*}
 \begin{tabular}{|c||c|c|c|c|c|c|c|c|c|}
\hline
$k_T$ [\GeVc] & $R_{o}$ [fm] & $R_{s}$ [fm] & $R_{l}$ [fm] & $\lambda$  & $M_1$ (\GeVc)$^{-2}$ & $M_2$ (\GeVc)$^{-2}$ & $M_3$ GeV$^{-2}$ & $M_4$ GeV$^{-1}$ & $\chi^2/{\rm ndf}$ \\
\hline
$[0.15,0.25]$ &  $1.06 \pm 0.03$& $1.00 \pm 0.04$& $1.38 \pm 0.05$& $0.665 \pm 0.005$ & \multirow{4}{*}{$0.43 \pm 0.07$} & \multirow{4}{*}{$0.22 \pm 0.06$} & \multirow{4}{*}{$1.51 \pm 0.12$} & \multirow{4}{*}{$1.02 \pm 0.09$} & \multirow{4}{*}{ 2218 / 336} \\
$[0.25,0.35]$ &  $0.96 \pm 0.02$& $0.95 \pm 0.03$& $1.21 \pm 0.03$& $0.588 \pm 0.006$ & & & & & \\
$[0.35,0.45]$ &  $0.89 \pm 0.02$& $0.88 \pm 0.02$& $1.08 \pm 0.04$& $0.579 \pm 0.009$ & & & & & \\
$[0.45,0.60]$ &  $0.78 \pm 0.04$& $0.79 \pm 0.02$& $0.94 \pm 0.03$& $0.671 \pm 0.028$ & & & & & \\
\hline
\end{tabular}
\caption{Fit results from a fit to data from \pp collisions at \roots = 200 GeV using Eq.~\ref{eq:cf3D} 
to parameterize the femtoscopic correlations and Eq.~\ref{eq:EMCICs} for non-femtoscopic ones 
(``EMCIC fit''). \label{tab:3DfitEMCICs}}
\end{table*}

\begin{figure}[ht!]
{\centerline{\includegraphics[width=0.48\textwidth]{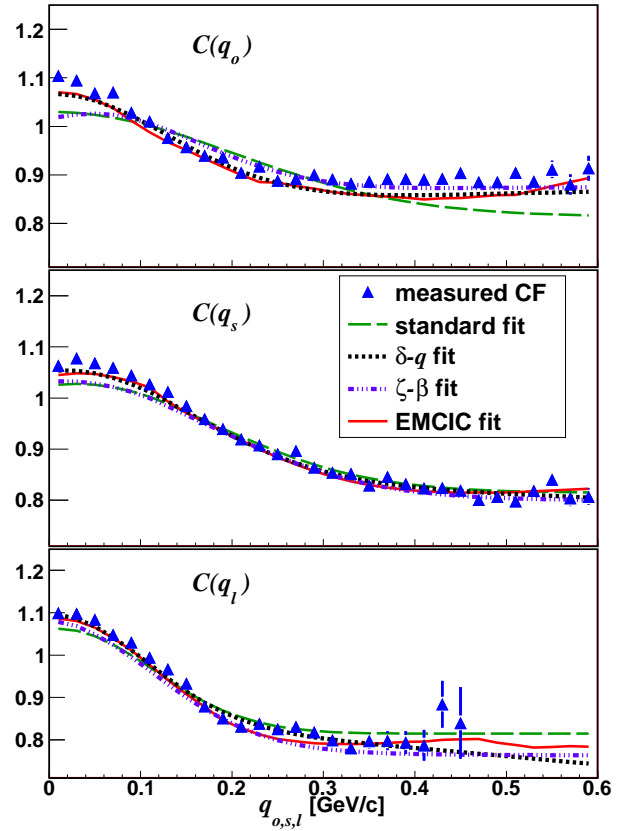}}}
\caption{(Color online) Cartesian projections of the 3D correlation function from \pp collisions 
at \roots=200 GeV for $k_T=[0.35,0.45]$ \GeVc~(blue triangles). 
Femtoscopic effects are parameterized with the form in Eq.~\ref{eq:cf3D};
different curves represent various parameterizations 
of non-femtoscopic correlations used in the fit and 
described in detail in Sec.~\ref{sec:nonfemto}.\label{fig:CartesianProjections}}
\end{figure}

\begin{figure}[ht!]
{\centerline{\includegraphics[width=0.5\textwidth]{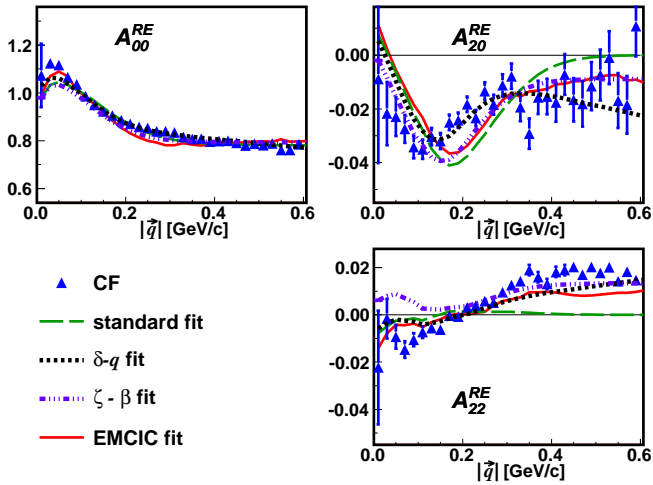}}}
\caption{(Color online) The first three non-vanishing moments of the spherical harmonic decomposition of 
the correlation function from \pp collisions at \roots=200 GeV, for $k_T = [0.15,0.25]$ \GeVc.
Femtoscopic effects are parameterized with the form in Eq.~\ref{eq:cf3D}; different curves represent various parameterizations 
of non-femtoscopic correlations used in the fit and 
described in detail in Sec.~\ref{sec:nonfemto}. The superscript ``RE'' in \mbox{$A^{RE}_{l,m}$}  stands for the real part of \Alm.\label{fig:SHDkt1}}
\end{figure}

\begin{figure}[ht!]
{\centerline{\includegraphics[width=0.5\textwidth]{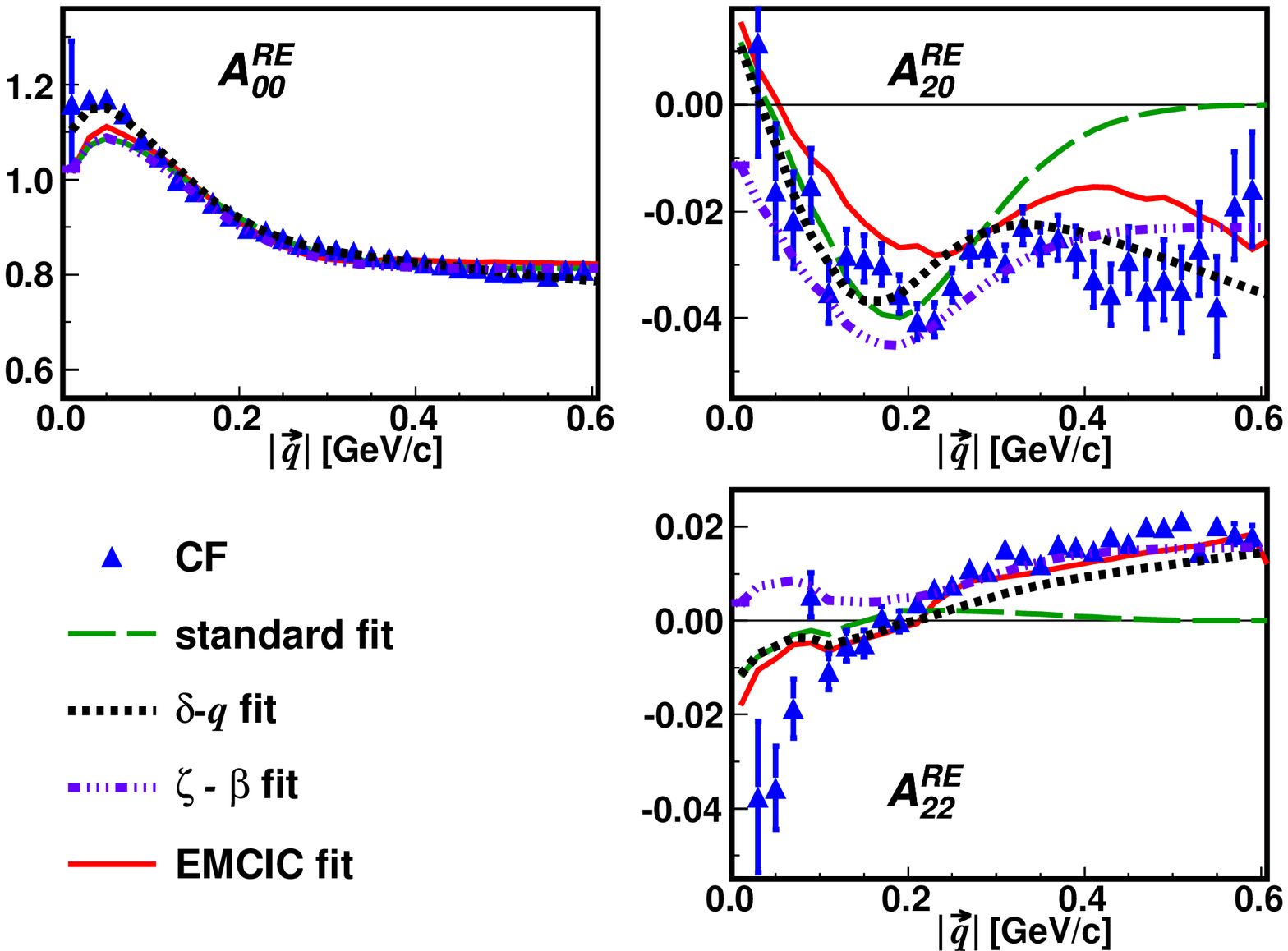}}}
\caption{(Color online) As for Fig.~\ref{fig:SHDkt1}, but for $k_T = [0.25,0.35]$ \GeVc.
\label{fig:SHDkt2}}
\end{figure}

\begin{figure}[t!]
{\centerline{\includegraphics[width=0.5\textwidth]{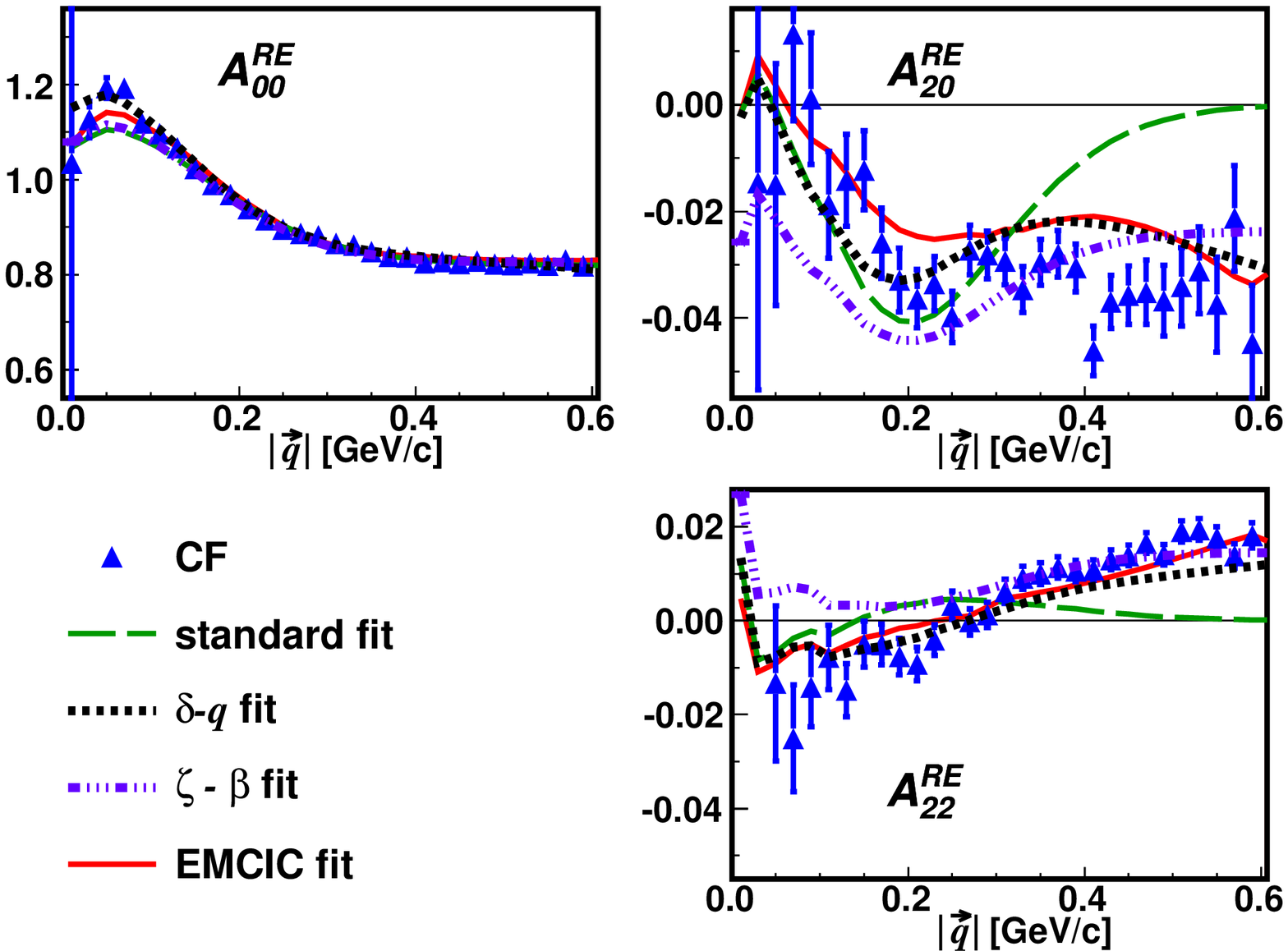}}}
\caption{(Color online) As for Fig.~\ref{fig:SHDkt1}, but for $k_T = [0.35,0.45]$ \GeVc.
\label{fig:SHDkt3}}
\end{figure}

\begin{figure}[ht!]
{\centerline{\includegraphics[width=0.5\textwidth]{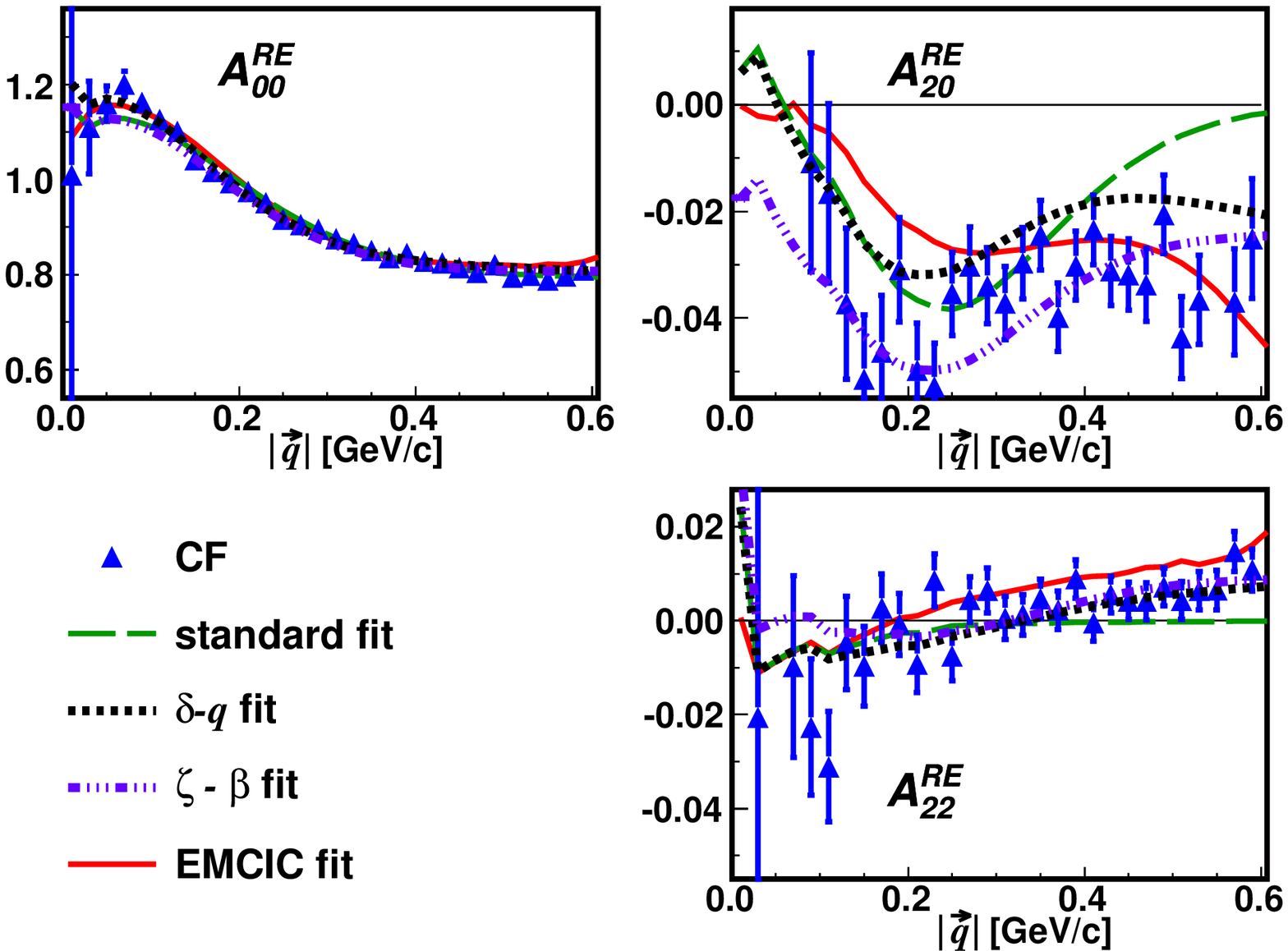}}}
\caption{(Color online) As for Fig.~\ref{fig:SHDkt1}, but for $k_T = [0.45,0.60]$ \GeVc.
\label{fig:SHDkt4}}
\end{figure}

Femtoscopic scales from three-dimensional correlation functions are usually extracted by fitting to the functional form
given in Equation~\ref{eq:cf3D}.  
In order to make connection to previous measurements, we employ the same form and vary
  the treatment of non-femtoscopic effects as discussed in Section~\ref{sec:nonfemto}.
The fits are shown as curves in Figs.~\ref{fig:CartesianProjections}-\ref{fig:SHDkt4}; the 
  slightly fluctuating structure observable in the sensitive spherical harmonic representation in Figs.~\ref{fig:SHDkt1}-\ref{fig:SHDkt4}
  results from finite-binning effects in plotting~\cite{Kisiel:2009iw}.

Dashed green curves in Figs.~\ref{fig:CartesianProjections}-\ref{fig:SHDkt4} represent the ``standard fit,'' in which non-femtoscopic
correlations are neglected altogether ($\Omega=1$).  Black dotted and purple dashed curves, respectively, indicate ``$\delta-q$''
(Equation~\ref{eq:dQ3D}) and ``$\zeta-\beta$'' (Equation~\ref{eq:ZetaBeta}) forms.  Solid red curves represent fits in which the
non-femtoscopic contributions follow the EMCIC (Equation~\ref{eq:EMCICs}) form.  None of the functional forms perfectly fits the
experimental correlation function, though the non-femtoscopic structure is semi-quantitatively reproduced by the ad-hoc $\delta-q$ and
$\zeta-\beta$ fits (by construction) and the EMCIC fit (non-trivially).  Rather than invent yet another ad-hoc functional form to better
fit the data, we will consider the radii produced by all of these forms.

The fit parameters for these four fits, for each of the four $k_T$ bins, are given in Tables~\ref{tab:3DfitStandard}-\ref{tab:3DfitEMCICs}.
Considering first the non-femtoscopic correlations, we observe that the ad-hoc fit parameters $\delta_{O,S,L}$ in Table~\ref{tab:3DfitDeltaQ}
and $\zeta$ and $\beta$ in Table~\ref{tab:3DfitZetaBeta} are different for each $k_T$ bin.  Due to their physical meaning, the EMCIC parameters $M_{1-4}$ are fixed for all $k_T$ values,
as indicated in Table~\ref{tab:3DfitEMCICs}.  Setting the characteristic particle mass to that of the pion and using Equations~\ref{eq:fitParameters}, \ref{eq:Esq}
and~\ref{eq:Nmin}, the non-femtoscopic parameters listed in Table~\ref{tab:3DfitEMCICs} correspond to the following values characteristic of the emitting system:
\begin{linenomath*}
\begin{align*}
N&=14.3 \pm  4.7 \\
\langle p_T^2 \rangle &= 0.17 \pm 0.06~{\rm (\GeVc)^2} \\
\langle p_z^2 \rangle &= 0.32 \pm 0.13~{\rm (\GeVc)^2}\\
\langle E^2 \rangle &= 0.51 \pm 0.11~{\rm GeV^2}\\
\langle E \rangle &= 0.68 \pm 0.08~{\rm GeV} .
\end{align*}
\end{linenomath*}
These values are rather reasonable~\cite{Chajecki:2008yi}.

\begin{figure}[ht!] 
{\centerline{\includegraphics[width=0.48\textwidth]{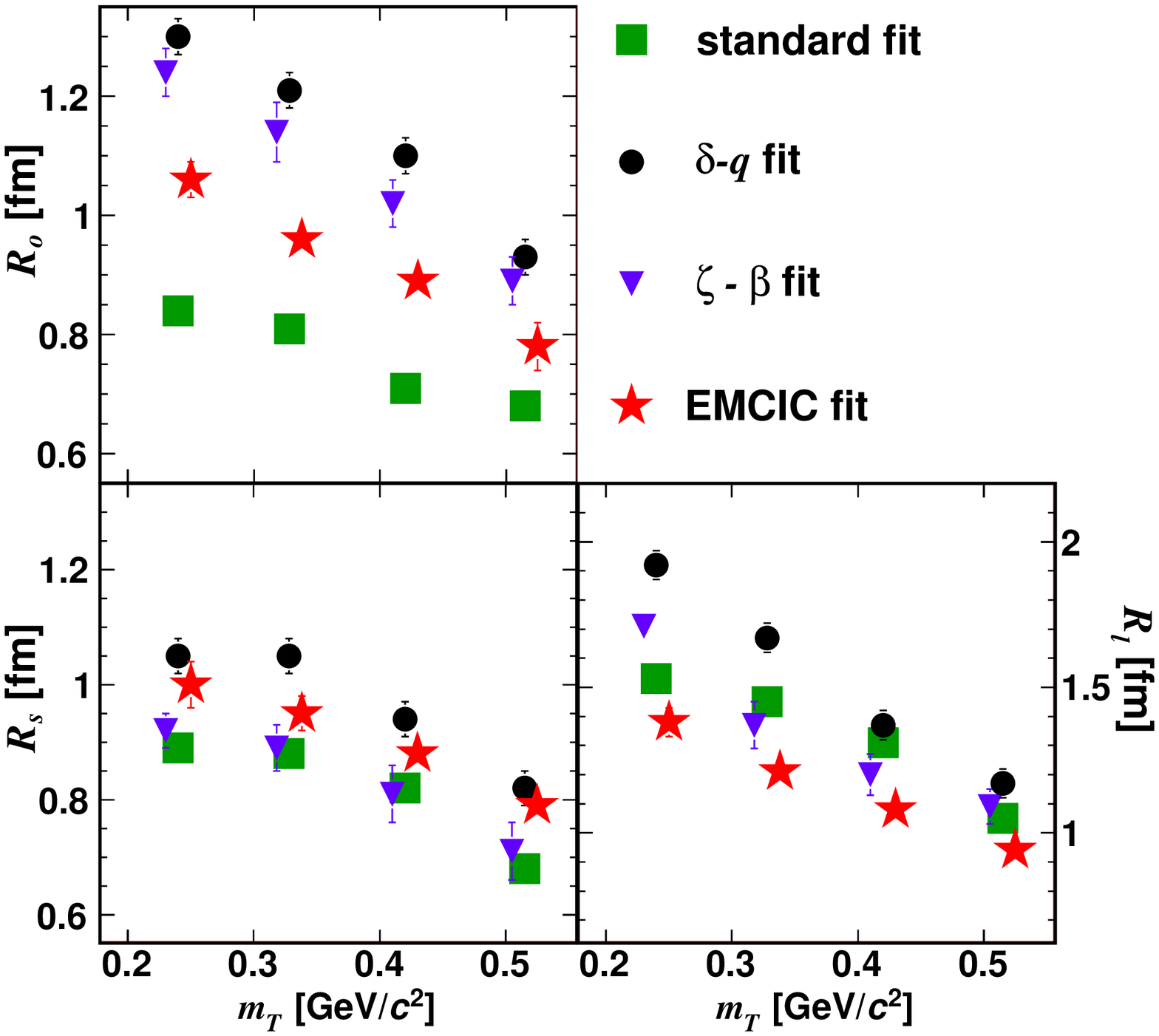}}}
\caption{(Color online) The $m_T$-dependence of the 3D femtoscopic radii in \pp collisions at \rootsTWO for different 
parameterizations of the non-femtoscopic correlations. See text for more details.  Data have been shifted slightly in the 
abscissa, for clarity.
\label{fig:mTdep3Dallfits}}
\end{figure}

HBT radii from the different fits are plotted as a function of transverse mass in Fig.~\ref{fig:mTdep3Dallfits}.
The treatment of the non-femtoscopic correlations significantly affects the magnitude of the femtoscopic length
scales extracted from the fit, especially in the ``out'' and ``long'' directions, for which variations up to 50\%
in magnitude are observed.  
The dependence of the radii on $m_T\equiv\sqrt{k_T^2+m^2}$ is quite similar in all cases.
We discuss this dependence further in Section~\ref{sec:WorldSystematics}.


\subsection{Transverse mass and multiplicity dependence of 1D femtoscopic radii}
\label{sec:mTandmultDep1D}

Since three-dimensional correlation functions encode
more information about the homogeneity region than do
one-dimensional correlation functions, they are also more statistics
hungry. Therefore, most previous particle physics experiments have
constructed and analyzed the latter.
For the sake of making the connection between our results and existing world systematics,
we perform similar analyses as those found in the literature.

The first important connection to make is for the $m_T$-dependence of HBT radii from minimum-bias \pp collisions.
We extract the one-dimensional HBT radius $R_{\rm inv}$ associated with the femtoscopic form in Equation~\ref{eq:CFqinvWithCoulomb},
using three forms for the non-femtoscopic terms.
For four selections in $k_T$, Table~\ref{tab:mTDepRinvSt} lists the fit parameters
for the ``standard'' fit that neglects non-femtoscopic correlations altogether ($\Omega=1$).
Tables~\ref{tab:mTDepRinvDeltaQ} and~\ref{tab:mTDepRinvEMCICs} list results when using the 1-dimensional $\delta-q$ form (Equation~\ref{eq:dQ1D})
and the EMCIC form (Equation~\ref{eq:EMCICs}), respectively.  In performing the EMCIC fit, the non-femtoscopic parameters $M_{1-4}$ were
kept fixed at the values listed in Table~\ref{tab:3DfitEMCICs}.

\begin{figure}[ht!]
{\centerline{\includegraphics[width=0.5\textwidth]{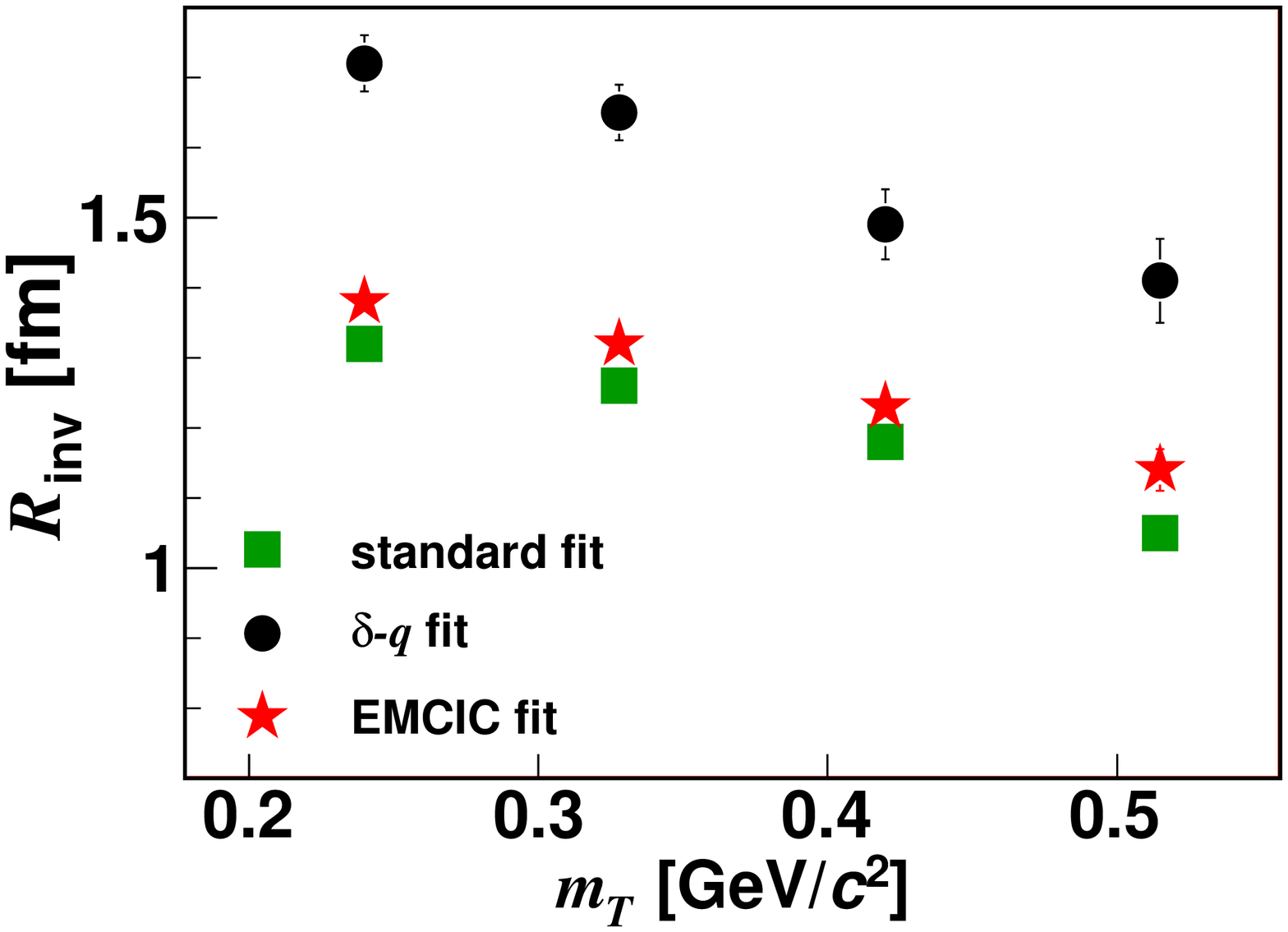}}}
\caption{(Color online) The $m_T$-dependence of $R_{\rm inv}$ from \pp collisions at \rootsTWO
for different parameterizations of the non-femtoscopic correlations used in the fit procedure. 
\label{fig:mTdep1Dallfits}}
\end{figure}

The one-dimensional radii from the three different treatments of non-femtoscopic effects are plotted as a function of $m_T$ in Fig.~\ref{fig:mTdep1Dallfits}.
The magnitude of the radius using the ad-hoc $\delta-q$ fit is $\sim 25\%$ larger than that from either the standard or EMCIC fit, but again all show similar
dependence on $m_T$.

%
\begin{table}
\begin{tabular}{|c|c|c|c|}
\hline
$k_T$ [\GeVc]  & $R_{\rm inv}$ [fm] & $\lambda$ & $\chi^2/{\rm ndf}$ \\
\hline
$[0.15,0.25]$ & $ 1.32 \pm  0.02 $ & $ 0.345 \pm 0.005  $ & 265 / 27 \\
$[0.25,0.35]$ & $ 1.26 \pm  0.02 $ & $ 0.357 \pm 0.007  $ & 203 / 27 \\
$[0.35,0.45]$ & $ 1.18 \pm  0.02 $ & $ 0.348 \pm 0.008  $ & 243 / 27 \\
$[0.45,0.60]$ & $ 1.05 \pm  0.03 $ & $ 0.413 \pm 0.012  $ & 222 / 27 \\
\hline
\end{tabular}
\caption{Fit results from a fit to 1D correlation function from \pp collisions at \rootsTWO using Eq.~\ref{eq:CFqinv} 
to parameterize the femtoscopic correlations (``standard fit''). \label{tab:mTDepRinvSt}}
\end{table}
\begin{table}
\begin{tabular}{|c|c|c|c|c|}
\hline
$k_T$ [\GeVc]  & $R_{\rm inv}$ [fm] & $\lambda$ & $\delta$ & $\chi^2/{\rm ndf}$ \\
\hline
$[0.15,0.25]$ & $ 1.72 \pm  0.04 $ & $ 0.285 \pm 0.007  $ & $ 0.237 \pm 0.007 $ & 86 / 26 \\
$[0.25,0.35]$ & $ 1.65 \pm  0.04 $ & $ 0.339 \pm 0.009  $ & $ 0.163 \pm 0.008 $ & 80 / 26 \\
$[0.35,0.45]$ & $ 1.49 \pm  0.05 $ & $ 0.308 \pm 0.011  $ & $ 0.180 \pm 0.015 $ & 71 / 26 \\
$[0.45,0.60]$ & $ 1.41 \pm  0.06 $ & $ 0.338 \pm 0.016  $ & $ 0.228 \pm 0.017 $ & 78 / 26 \\
\hline
\end{tabular}
\caption{Fit results from a fit to 1D correlation function from \pp collisions at \rootsTWO using Eq.~\ref{eq:CFqinv} 
to parameterize the femtoscopic correlations  and Eq.~\ref{eq:dQ1D} for non-femtoscopic ones
(``$\delta-q$ fit''). \label{tab:mTDepRinvDeltaQ}}
\end{table}
\begin{table}
\begin{tabular}{|c|c|c|c|}
\hline
$k_T$ [\GeVc]  & $R_{\rm inv}$ [fm] & $\lambda$ & $\chi^2/{\rm ndf}$ \\
\hline
$[0.15,0.25]$ & $ 1.38 \pm  0.03 $ & $ 0.347 \pm 0.005  $ & 99 / 27 \\
$[0.25,0.35]$ & $ 1.32 \pm  0.03 $ & $ 0.354 \pm 0.006  $ & 97 / 27 \\
$[0.35,0.45]$ & $ 1.23 \pm  0.04 $ & $ 0.349 \pm 0.009  $ & 86 / 27 \\
$[0.45,0.60]$ & $ 1.14 \pm  0.05 $ & $ 0.411 \pm 0.013  $ & 80 / 27 \\
\hline
\end{tabular}
\caption{Fit results from a fit to 1D correlation function from \pp collisions at \rootsTWO using Eq.~\ref{eq:CFqinv} 
to parameterize the femtoscopic correlations  and Eq.~\ref{eq:EMCICs} for non-femtoscopic ones
(``EMCIC fit''). The non-femtoscopic parameters $M_{1-4}$ were not varied, but kept fixed
to the values in Table~\ref{tab:3DfitEMCICs}.
\label{tab:mTDepRinvEMCICs}}  
\end{table}

In order to compare with the multiplicity dependence of $k_T$-integrated HBT radii reported in high energy particle collisions,
  we combine $k_T$ bins and separately analyze low ($dN_{ch}/d\eta\leq 6$) and high ($dN_{ch}/d\eta\geq 7$) multiplicity events.
The choice of the cut was dictated by the requirement of sufficient pair statistics in the two event classes.
Fit parameters for common fitting functions are given in Table~\ref{tab:multDep1D}, for minimum-bias and multiplicity-selected collisions.

\begin{figure}[ht!]
{\centerline{\includegraphics[width=0.5\textwidth]{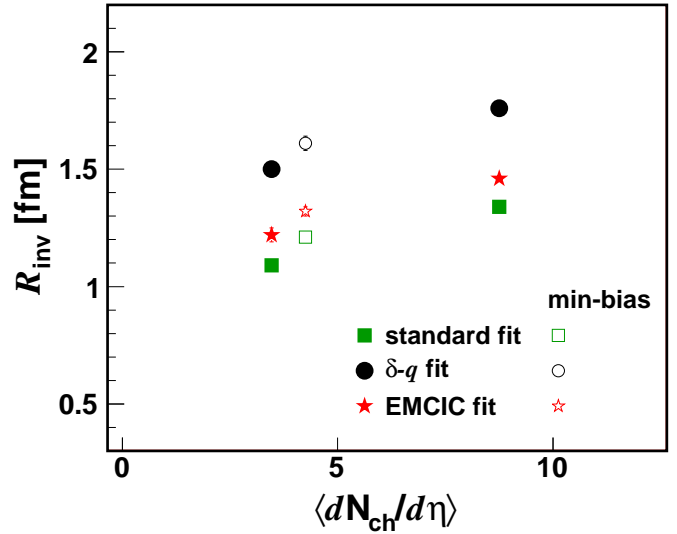}}}
\caption{(Color online) The multiplicity dependence of $R_{\rm inv}$ from \pp collisions at \rootsTWO for different 
parameterizations of the non-femtoscopic correlations. Pions within the range of $k_T=[0.15,0.60]$ \GeVc~were used in the analysis. \label{fig:multDepRinv}}
\end{figure}

Figure~\ref{fig:multDepRinv} shows the multiplicity dependence of the common one-dimensional HBT radius $R_{\rm inv}$, extracted by 
  parameterizing the femtoscopic correlations according to Equation~\ref{eq:CFqinvWithCoulomb}.
Non-femtoscopic effects were either ignored (``standard fit'' $\Omega=1$) or parameterized with the ``$\delta-q$'' (Eq.~\ref{eq:dQ1D})
  or EMCIC (Eq.~\ref{eq:EMCICs}) functional form.
In order to keep the parameter count down, the EMCIC kinematic parameters ($\langle p_T^2\rangle$, $\langle p_z^2\rangle$, $\langle E^2\rangle$,
   $\langle E\rangle$) were kept fixed to the values obtained from the 3-dimensional fit, and only $N$ was allowed to vary.
In all cases, $R_{\rm inv}$ is observed to increase with multiplicity.  Parameterizing non-femtoscopic effects according to the
  EMCIC form gives similar results as a ``standard'' fit ignoring them, whereas the ``$\delta-q$'' form generates an offset of approximately
  0.3 fm, similar to all three- and one-dimensional fits discussed above.
That different numerical values are obtained for somewhat different fitting functions, is not surprising.
The point we focus on is that the systematic dependences of the femtoscopic scales, both with $k_T$ and multiplicity, are robust.

Table~\ref{tab:multDep2D} lists fit parameters to two-dimensional correlation functions in $q_T$ and $q_0$, using Equations~\ref{eq:CFcp} and~\ref{eq:CFgauss}. 
The radius from the former fit is approximately twice that of the latter, as expected (c.f. Sec.~\ref{sec:femto}).
These values will be compared with previously measured data in the next section.

\begin{table*}
\begin{tabular}{|c|c||c||c|c|}
\hline
\multirow{2}{*}{method} & \multirow{2}{*}{fit parameter} & \multicolumn{3}{|c|}{$\langle dN_{ch}/d\eta \rangle$} \\
\cline{3-5}
        &       & $ 4.25$ (min-bias) & $ 3.47 $ & $ 8.75 $ \\
\hline
\multirow{3}{*}{standard fit}             & $R_{\rm inv}$ [fm]  & $ 1.21 \pm 0.01 $   & $ 1.09 \pm 0.02 $  & $ 1.34 \pm 0.02 $  \\
                                          & $\lambda$           & $ 0.353 \pm 0.003 $ & $ 0.347 \pm 0.04 $ & $ 0.356 \pm 0.03 $ \\
                                          & $\chi^2/{\rm ndf}$  & 202 / 27            & 100 / 27           & 92 / 27            \\
\hline
\multirow{4}{*}{$\delta-q$ fit}  & $R_{\rm inv}$ [fm]           & $ 1.61 \pm 0.01 $   & $ 1.50 \pm 0.03 $   & $ 1.76 \pm 0.03 $   \\
                                 & $\lambda$                    & $ 0.312 \pm 0.003 $ & $ 0.275 \pm 0.005 $ & $ 0.322 \pm 0.007 $ \\
                                 & $\delta Q_{\rm inv}$ [\cGeV] & $-0.191 \pm 0.003 $ & $-0.242 \pm 0.005 $ & $-0.194 \pm 0.006 $ \\
                                 & $\chi^2/{\rm ndf}$           & 159 / 26            & 83 / 26             & 73 / 26             \\
\hline
\multirow{4}{*}{EMCIC fit}       & $R_{\rm inv}$ [fm]           & $ 1.32 \pm 0.02 $   & $ 1.22 \pm 0.03 $   & $ 1.46 \pm 0.02 $   \\
                                 & $\lambda$                    & $ 0.481 \pm 0.003 $ & $ 0.485 \pm 0.003 $ & $ 0.504 \pm 0.004 $ \\
                                 & $N$                          & $ 14.3 \pm 4.7  $   & $ 11.8 \pm 7.1 $    & $ 26.3 \pm 8.4 $    \\
                                 & $\chi^2/{\rm ndf}$           & 161 / 26            & 80 / 26             & 75 / 26             \\
\hline

\end{tabular}
\caption{Multiplicity dependence of fit results to 1D correlation function from \pp collisions at \rootsTWO
for different fit parameterizations. \label{tab:multDep1D}}
\end{table*}


\begin{table*}                                                                                                                                                                
\begin{tabular}{|c|c||c||c|c|}                                                                                                                                          
\hline                                                                                                                                                                    
\multirow{2}{*}{method} & \multirow{2}{*}{fit parameter} &
\multicolumn{3}{|c|}{$\langle dN_{ch}/d\eta \rangle$} \\  
\cline{3-5}       &       & $ 4.25$ (min-bias) & $ 3.47 $ & $ 8.75 $ \\                                                                                                                
\hline
\multirow{4}{*}{Eq.~\ref{eq:CFcp}} & $R_{B}$ [fm]       & $ 1.79 \pm 0.01 $ & $ 1.61 \pm 0.02 $ & $ 1.92 \pm 0.02 $ \\ 
                                   & $\tau$  [fm/c]     & $ 1.03 \pm 0.02 $ & $ 0.98 \pm 0.02 $ & $ 1.24 \pm 0.03 $ \\ 
                                   & $\lambda$          & $0.353\pm 0.003 $ & $0.354\pm 0.003 $ & $0.334\pm 0.004 $ \\
                                   & $\chi^2/{\rm ndf}$ & 5308 / 896        & 2852 / 896        & 1890 / 896        \\
\hline                                                                                                                                                                        
\multirow{4}{*}{Eq.~\ref{eq:CFgauss}} & $R_{G}$ [fm]       & $ 1.01 \pm0.01 $  & $ 0.89 \pm 0.01 $ & $ 1.07 \pm 0.01 $ \\  
                                      & $\tau$ [fm/c]      & $ 0.76 \pm 0.01 $ & $ 0.73 \pm 0.02 $ & $ 0.91 \pm 0.02 $ \\
                                      & $\lambda$          & $0.353\pm 0.003 $ & $0.352\pm 0.003 $ & $0.332\pm 0.004 $ \\
                                      & $\chi^2/{\rm ndf}$ & 5749 / 896        & 3040 / 896        & 2476 / 896        \\
\hline                                                                                                                                                                        
\end{tabular}                                                                                                                                                                 
\caption{Multiplicity dependence of fit parameters to two-dimensional correlation functions
from \pp collisions at \rootsTWO using Equations~\ref{eq:CFcp} and~\ref{eq:CFgauss}.
To consistently compare to previous measurements, $\Omega$ was set to unity (c.f. Equation~\ref{eq:factorization}).
\label{tab:multDep2D}}
\end{table*}


\section{Comparison with world systematics}
\label{sec:WorldSystematics}

In this section, we make the connection between femtoscopic measurements in heavy ion collisions and those in particle physics, by
  placing our results in the context of world systematics from each.

\subsection{Results in the Context of Heavy Ion Systematics}
\label{sec:HIsystematics}

The present measurements represent the first opportunity to study femtoscopic correlations from hadronic collisions
  and heavy ion collisions, using the same detector, reconstruction, analysis and fitting techniques.
The comparison should be direct, and differences in the extracted HBT radii should arise from differences in the source geometry itself.
In fact, especially in recent years, the heavy ion community has generally arrived at a consensus among the different experiments,
  as far as analysis techniques, fitting functions and reference frames to use.
This, together with good documentation of event selection and acceptance cuts, has led to a quantitatively consistent world
  systematics of femtoscopic measurements in heavy ion collisions over two orders of magnitude in collision energy~\cite{Lisa:2005dd};
  indeed, at RHIC, the agreement in HBT radii from the different experiments is remarkably good.
Thus, inasmuch as STAR's measurement of HBT radii from \pp collisions may be directly compared with STAR's HBT radii from \AuAu collisions,
  they may be equally well compared to the world's systematics of all heavy ion collisions.

\begin{figure}[ht!]
{\centerline{\includegraphics[width=0.5\textwidth]{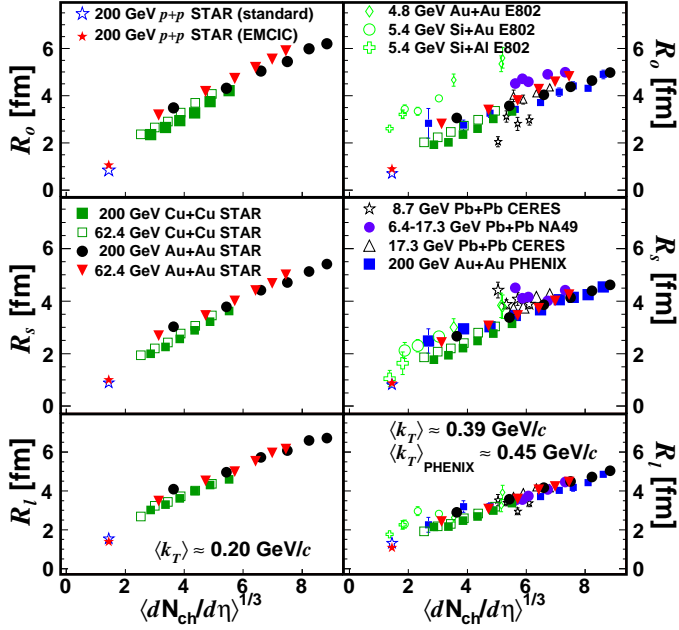}}}
\caption{(Color online) The multiplicity dependence of the HBT radii from \pp, \CuCu~\cite{Abelev:2009tp} and \AuAu~\cite{Adams:2004yc,Abelev:2009tp} 
collisions from STAR compared with results from other experiments~\cite{Lisa:2005dd}.  Left and right panels show radii measured with $\langle k_T\rangle \approx 0.2$
and 0.39~\GeVc, respectively.  Radii from \pp collisions are shown by blue (``standard fit'') and red (``EMCIC fit'') stars.
\label{fig:multScaling}}
\end{figure}

As with most heavy ion observables at low transverse momentum~\cite{Caines:2006if}, the HBT radii $R_{s}$ and $R_{l}$ scale primarily with event multiplicity~\cite{Lisa:2005dd}
  (or, at lower energies, with the number of particles of different species~\cite{Adamova:2002ff,Lisa:2008gf}) rather than with energy or impact parameter.
The radius $R_{o}$, which nontrivially combines space and time, shows a less clear scaling~\cite{Lisa:2005dd}, retaining some energy dependence.
As seen in Fig.~\ref{fig:multScaling}, the radii from \pp collisions at \rootsTWO fall naturally in line with this multiplicity scaling.
On the scale relevant for this comparison, the specific treatment of non-femtoscopic correlations is unimportant.

\begin{figure}[ht!]
{\centerline{\includegraphics[width=0.5\textwidth]{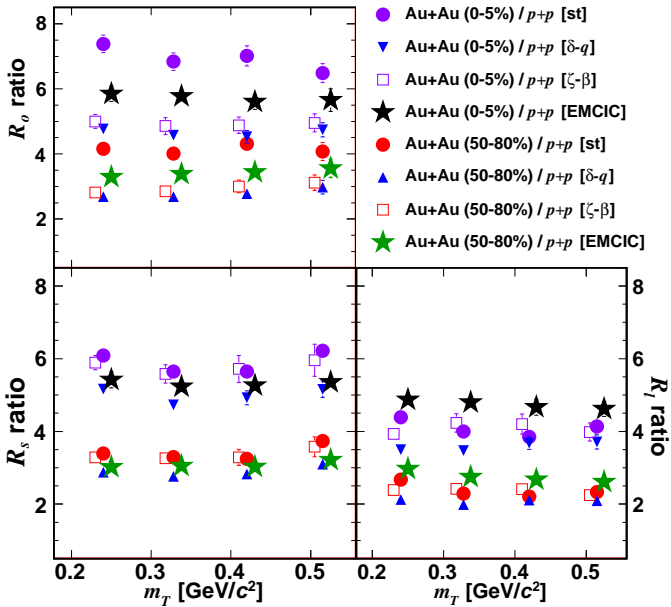}}}
\caption{(Color online) The ratio of the HBT radii from \AuAu collisions~\cite{Adams:2004yc} to results from \pp collisions plotted versus the transverse mass.\label{fig:mTscaling}}
\end{figure}

One of the most important systematics in heavy ion femtoscopy is the $m_T$-dependence of HBT radii, which directly measures space-momentum correlations
  in the emitting source at freeze-out; in these large systems, the $m_T$-dependence is often attributed to collective flow~\cite{Retiere:2003kf}.
As we saw in Fig.~\ref{fig:mTdep3Dallfits}, a significant dependence is seen also for \pp collisions.
Several authors~\citep[e.g.][]{Alexopoulos:1992iv,Alexopoulos:2002eh,Agababyan:1996rg,collaboration:2007he,Kittel:2001zw}
  have remarked on the qualitative ``similarity'' of the $m_T$-dependence of HBT radii measured in high energy particle collisions, but the
  first direct comparison is shown in Fig.~\ref{fig:mTscaling}.
There, the ratios of the three dimensional radii 
  in \AuAu collisions to \pp radii obtained with different treatments of the non-femtoscopic correlations,
  are plotted versus $m_T$.
Well beyond qualitative similarity, the ratios are remarkably flat-- i.e. the $m_T$-dependence in \pp collisions is
  quantitatively almost identical to that in \AuAu collisions at RHIC.
We speculate on the possible meaning of this in Section~\ref{sec:WorldSys}.

\subsection{Results in the context of high-energy particle measurements}
\label{sec:WorldSys}

Recently, a review of the femtoscopic results~\cite{Chajecki:2009zg}
 from particle collisions like \pp, \ppbar and                   
 \epem studied at different energies has been published. Here, we compare 
 STAR results from \pp collisions at \rootsTWO with world systematics.

\begin{figure}[ht!]
{\centerline{\includegraphics[width=0.5\textwidth]{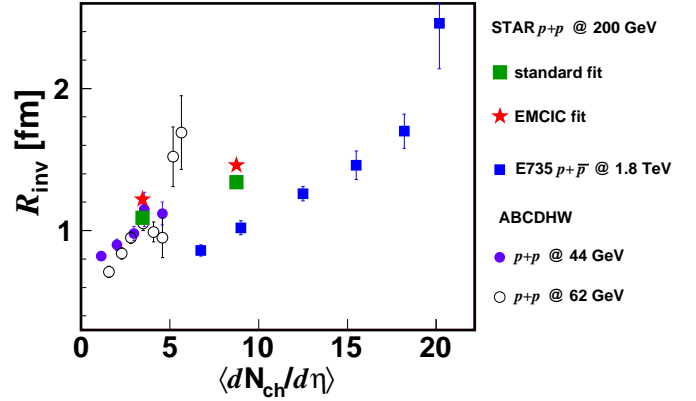}}}
\caption{(Color online) The multiplicity dependence of the 1D femtoscopic radius $R_{\rm inv}$ from hadronic collisions
measured by STAR,
 E735~\cite{Alexopoulos:1992iv}, 
and ABCDHW~\cite{Breakstone:1986xs} collaborations.
\label{fig:wsmultdep_1Donly}}
\end{figure}

\begin{figure}[ht!]
{\centerline{\includegraphics[width=0.5\textwidth]{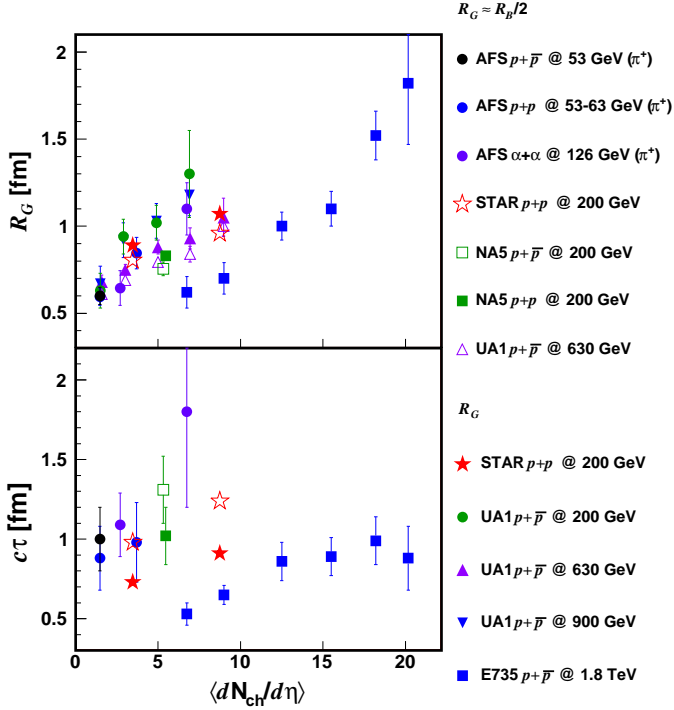}}}
\caption{(Color online) The multiplicity dependence of radius and timescale parameters to 2-dimensional correlation
functions 
measured by STAR,
 E735~\cite{Alexopoulos:1992iv},
 UA1~\cite{Albajar:1989sj},
 AFS~\cite{Akesson:1983gk} and
 NA5~\cite{DeMarzo:1984zr}.
The legend on the right indicates that the first 7 sets of datapoints come from fits to Eq.~\ref{eq:CFcp}, in
which case the parameter $R_B/2$ is plotted in the upper panel; the last 5 sets of datapoints come from fits to
Eq.~\ref{eq:CFgauss}, for which $R_G$ is plotted.  As discussed in Section~\ref{sec:femto} and confirmed by STAR
and UA1, $R_G\approx R_B/2$.
The UA1 Collaboration set $\tau\equiv 0$ in their fits.
\label{fig:wsmultdep_2Donly}}
\end{figure}

The multiplicity dependence of femtoscopic parameters from one- and two-dimensional correlation functions are shown in Figs.~\ref{fig:wsmultdep_1Donly} and~\ref{fig:wsmultdep_2Donly}.
For any given experiment, the radius parameter increases with event multiplicity.
However, in contrast to the nearly ``universal'' multiplicity dependence seen in heavy ion collisions (c.f. Fig.~\ref{fig:multScaling}), only a qualitative trend is
  observed, when the different measurements are compared.

There are several possible reasons for this lack of ``universality''~\cite{Chajecki:2009zg}.
Clearly one possibility is that there is no universal multiplicity dependence of the femtoscopic scales;
  the underlying physics driving the space-time freezeout geometry may be quite different, considering
  \roots~ varies from 44 to 1800~GeV in the plot.
However, even if there were an underlying universality between these systems, it is not at all clear that it would appear in this figure,
  due to various difficulties in tabulating historical data~\cite{Chajecki:2009zg}.
Firstly, as discussed in Section~\ref{sec:formalism} the experiments used different fitting functions to extract the HBT radii, making
  direct comparison between them difficult.  
Secondly, as we have shown, the radii depend on both multiplicity and $k_T$.  
Since, for statistical reasons, the results in Fig.~\ref{fig:multScaling}
  are integrated over the acceptance of each experiment, and these acceptances differ strongly, any universal scaling would be obscured.  
For example, since the acceptance of Tevatron experiment E735~\cite{Alexopoulos:1992iv} is weighted towards higher $k_T$ than the other
  measurements, one expects a systematically lower HBT radius, at a given multiplicity.
Indeed, even the ``universal'' multiplicity scaling in heavy ion collisions is only universal for a fixed selection in $k_T$.
Thirdly, the measure used to quantify the event multiplicity varies significantly in the historical literature;
  thus the determination 
  of $\langle dN_{\rm ch}/d\eta\rangle$ for any given experiment shown in Fig.~\ref{fig:multScaling} is only approximate.

From the discussion above, we cannot conclude definitively that there is-- or is not-- a universal multiplicity scaling of femtoscopic radii in 
  high energy hadron-hadron collisions.
We conclude only that an increase of these radii with multiplicity is observed in all measurements for which
  $\sqrt{s}\gtrsim 40$~GeV and that the present analysis of \pp collisions is consistent with world systematics.

\begin{figure}[t!]
\includegraphics[width=0.45\textwidth]{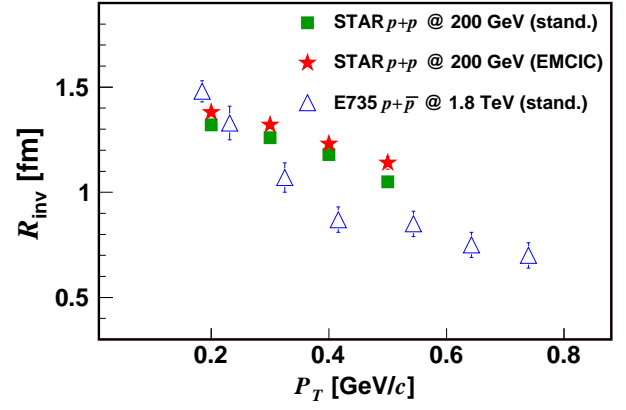}
\caption{(Color online) 
One-dimensional femtoscopic radii from \pp collisions at RHIC and \ppbar collisions at the Tevatron~\cite{Alexopoulos:1992iv}
are plotted versus the transverse momentum $P_T\equiv\left(\vec{p}_{1,T}+\vec{p}_{2,T}\right)/2$ (c.f. Eq.~\ref{eq:CFexp}).
\label{fig:wsmTdep_1Donly}}
\end{figure}

\begin{figure}[t!]
\includegraphics[width=0.45\textwidth]{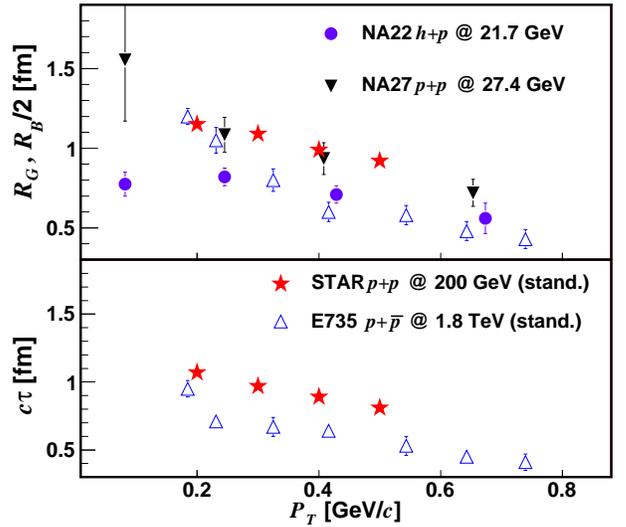}
\caption{(Color online) The transverse momentum dependence of fit parameters to two-dimensional correlation functions.
STAR results from fit to Equation~\ref{eq:CFgauss}, compared to measurements by
E735~\cite{Alexopoulos:1992iv}, 
 NA27~\cite{AguilarBenitez:1991ri}
and NA22~\cite{Agababyan:1993aj}.                    
The SPS experiments NA22 and NA27 set $\tau\equiv 0$ in their fits.  
STAR and E735 data plotted versus $P_T\equiv\left(\vec{p}_{1,T}+\vec{p}_{2,T}\right)/2$ (c.f. Eq.~\ref{eq:CFexp}).
NA27 reported results in terms of $|\vec{P}|$ and NA22 in terms of $2|\vec{P}|$.
For purposes of plotting here, $P_T = \sqrt(2/3)|\vec{P}|$ was assumed.
\label{fig:wsmTdep_2Donly}}
\end{figure}

In Section~\ref{sec:results}, we discussed the $p_T$-dependence of HBT radii observed in our analysis.
Previous experiments on high-energy collisions between hadrons-- and even leptons-- have reported similar trends.
As discussed above, direct comparisons with historical high-energy measurements are problematic.
Comparisons between fit parameters to 1- and 2-dimensional correlation functions are shown 
in Figs.~\ref{fig:wsmTdep_1Donly} and~\ref{fig:wsmTdep_2Donly}.
All experiments observe a decrease in femtoscopic parameters with increasing transverse momentum.  Our radii
at \roots=200~GeV fall off similarly or somewhat more than those measured at an order of magnitude lower energy
at the SPS~\cite{AguilarBenitez:1991ri,Agababyan:1996rg}, and less than those measured at an order of magnitude
higher energy at the Tevatron~\cite{Alexopoulos:1992iv}.
It is tempting to infer that this compilation indicates an energy evolution of the $p_T$-dependence of femtoscopic radii.
However, given our previous discussion, we conclude only that there is qualitative agreement between experiments
at vastly different collision energies, and all show similar $p_T$ dependence.

\begin{figure}[ht!]
{\centerline{\includegraphics[width=0.5\textwidth]{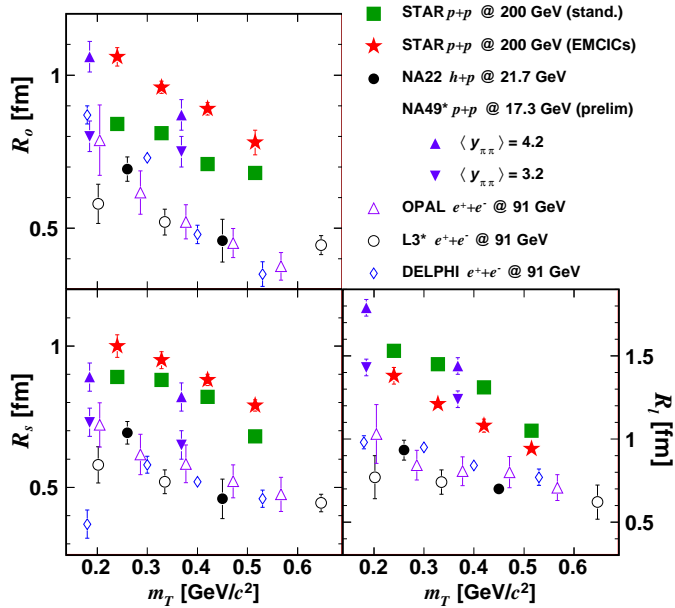}}}
\caption{(Color online) The transverse mass dependence of 3D femtoscopic radii from particle collisions.
Data from NA22~\cite{Agababyan:1996rg},
 NA49 preliminary~\cite{Ganz:1999ht},
 OPAL~\cite{collaboration:2007he},
 L3~\cite{Achard:2001za},
 DELPHI~\cite{Smirnova:1999aa}.
\label{fig:wsmTdep3D}}
\end{figure}

Systematics in 3-dimensional HBT radii from hadron collisions are less clear and less abundant, though our
  measurements are again qualitatively similar to those reported at the SPS, as shown in Fig.~\ref{fig:wsmTdep3D}.
There, we also plot recent results from \epem collisions at LEP; in those 3-dimensional analyses, the
  ``longitudinal'' direction is the thrust axis, whereas the beam axis is used in hadron-hadron collisions, as in
  heavy ion collisions.

\section{Discussion}
\label{sec:discussion}

We have seen that HBT radii from \pp collisions at RHIC are qualitatively consistent with the trends observed
  in particle collisions over a variety of collision energies.
Further, they fall quantitatively into the much better-defined world systematics for heavy ion collisions at RHIC
  and similar energies.
Particularly intriguing is the nearly identical dependence on $m_T$ of the HBT radii in \pp and heavy ion collisions,
  as this dependence is supposed~\cite{Pratt:1984su,Heinz:1999rw} to reflect the underlying dynamics of the latter.
Several possible sources of an $m_T$ dependence of HBT radii in small systems have been put forward to
  explain previous measurements.

1.
Alexander {\it et al.}~\cite{Alexander:1999rm,Alexander:2001gk} have suggested 
  that the Heisenberg uncertainty principle can produce the transverse momentum
  dependence of femtoscopic radii in \epem collisions.
However, as discussed in ~\cite{Chajecki:2009zg}, a more detailed study of the results from \epem collisions 
  complicates the quantitative comparisons of the data from various experiments 
  and thus the interpretation.
Additionally, the arguments from~\cite{Alexander:1999rm,Alexander:2001gk} apply only to the longitudinal direction ($R_{l}$), so could not explain
  the dependence of all three radii.

2.
In principle, string fragmentation should also generate space-momentum correlations 
  in small systems, hence an $m_T$ dependence of the HBT radii.
However, there are almost no quantitative predictions that can be compared with data.
The numerical implementation {\tt PYTHIA}, which incorporates the Lund string model into the soft sector dynamics,
  implements Bose-Einstein enhancement only as a crude parameterization designed to mock up the effect~\citep[c.f. Section 12.4.3 of][]{Sjostrand:2006za}
  for the purpose of estimating distortions to $W$-boson invariant mass spectrum.
Any Bose-Einstein correlation function may be dialed into the model, with 13 parameters to set the
  HBT radius, lambda parameter, and correlation shape; there is no first-principles predictive power.
On more general grounds, the mass dependence of the femtoscopic radii cannot be explained within 
  a Lund string model~\cite{Bialas:2000yi,Alexander:2001zx,Alexander:2003ca}.

3.
Long-lived resonances may also generate the space-momentum dependence of femtoscopic radii~\cite{Wiedemann:1996ig}.
However, as discussed in~\cite{Chajecki:2009zg}, the resonances would affect the HBT radii from \pp collisions differently 
  than those from \AuAu collisions, since the scale of the resonance ``halo'' is fixed by resonance lifetimes 
  while the scale of the ``core'' is different for the two cases.
Thus it would have to be a coincidence that the same $m_T$ dependence is observed in both systems.
Nevertheless, this avenue should be explored further.

4.
Bia{\l}as {\it et al.} have introduced a model~\cite{Bialas:2000yi} based on a direct proportionality
  between the four-momentum and space-time freeze-out position; this model successfully described data from \epem collisions.
The physical scenario is based on freezeout of particles emitted from a common tube, after a fixed time of 1.5~fm/c.
With a very similar model, Humanic~\cite{Humanic:2006ib} was able to reproduce femtoscopic radii measured
  at the Tevatron~\cite{Alexopoulos:1992iv} only with strong additional hadronic rescattering effects.
With rescattering in the final state, both the multiplicity- and the $m_T$-dependence of the radii were reproduced~\cite{Humanic:2006ib}.

5.
It has been suggested~\cite{Alexopoulos:1992iv,Agababyan:1996rg,Kittel:2001zw,Csorgo:2004id,collaboration:2007he}
  that the $p_T$-dependence of HBT radii in very small systems might reflect bulk collective flow, as it is believed to do in heavy ion collisions.
This is the only explanation that would automatically account for the nearly identical $p_T$-scaling discussed in
  Section~\ref{sec:HIsystematics}.
However, it is widely believed that the system created in \pp collisions is too small to generate bulk flow.

The remarkable similarity between the femtoscopic systematics in heavy ion and hadron collisions may well be coincidental.
Given the importance of the $m_T$-dependence of HBT radii in heavy ion collisions, and the unclear origin of this dependence in hadron collisions,
  further theoretical investigation is clearly called for.
Additional comparative studies of other soft-sector observables (e.g. spectra) may shed further light onto this coincidence.

\section{Summary}
\label{sec:summary}

We have presented a systematic femtoscopic analysis of two-pion correlation functions from \pp collisions at RHIC.
In addition to femtoscopic effects, the data show correlations due to energy and momentum conservation. 
Such effects have been observed previously in low-multiplicity measurements at Tevatron, SPS, and elsewhere. 
In order to compare to historical data and to identify systematic effects on the HBT radii, we have treated these effects
  with a variety of empirical and physically-motivated formulations.
While the overall magnitude of the geometric scales vary with the method, the important systematics do not.

In particular, we observe a significant positive correlation between the one- and three-dimensional radii and the multiplicity of the collision,
while the radii decrease with increasing transverse momentum.
Qualitatively, similar multiplicity and momentum 
systematics have been observed previously in measurements of hadron and electron collisions at the Sp$\overline{\rm p}$S, Tevatron, ISR 
and LEP. However, the results from these experiments could not be directly compared to those from heavy ion collisions, due to differences
in techniques, fitting methods, and acceptance.

Thus, the results presented here provide a unique possibility for a direct comparison of femtoscopy in \pp and \ApA collisions. 
 We have seen very similar $p_T$ and multiplicity scaling of the femtoscopic scales in \pp as in \ApA collisions, independent of the fitting 
  method employed.
Given the importance of femtoscopic systematics in understanding the bulk sector in \AuAu collisions, further exploration of the physics behind the same scalings in \pp collisions
  is clearly important, to understand our ``reference'' system.
The similarities observed could indicate a deep connection between the underlying
  physics of systems with size on order of the confinement scale, and of systems much larger.
Similar comparisons will be possible at the Large Hadron Collider, where the higher collision energies will render conservation
laws less important, especially for selections on the very highest-multiplicity collisions.

\section*{Acknowledgements}
We thank the RHIC Operations Group and RCF at BNL, the NERSC Center at LBNL and the 
Open Science Grid consortium for providing resources and support. This work was 
supported in part by the Offices of NP and HEP within the U.S. DOE Office of Science, 
the U.S. NSF, the Sloan Foundation, the DFG cluster of excellence `Origin and Structure 
of the Universe' of Germany, CNRS/IN2P3, STFC and EPSRC of the United Kingdom, FAPESP CNPq of 
Brazil, Ministry of Ed. and Sci. of the Russian Federation, NNSFC, CAS, MoST, and 
MoE of China, GA and MSMT of the Czech Republic, FOM and NWO of the Netherlands, DAE, 
DST, and CSIR of India, Polish Ministry of Sci. and Higher Ed., Korea Research Foundation, 
Ministry of Sci., Ed. and Sports of the Rep. Of Croatia, 
and RosAtom of Russia.

\end{document}